\documentclass[a4paper,fleqn]{cas-sc}

\usepackage[utf8]{inputenc}
\usepackage[english]{babel}
\usepackage[square,numbers,sort&compress]{natbib}

\usepackage{placeins}
\usepackage{amsmath}
\usepackage{graphicx}
\usepackage{tcolorbox}
\usepackage{relsize}
\usepackage{comment}
\usepackage{tabu,multirow}
\usepackage{epsfig}
\usepackage{array}
\usepackage{amssymb}
\usepackage{latexsym}
\usepackage{xcolor}
\usepackage{subcaption}
\usepackage{soul}
\usepackage[export]{adjustbox}
\usepackage{setspace}
\usepackage{caption}
\usepackage{float}
\usepackage{tikz}
\usetikzlibrary{positioning, shapes, arrows}
\usepackage{epsfig}
\usepackage{siunitx}
\usepackage{miller}
\usepackage[noabbrev,capitalise]{cleveref}
\usepackage{siunitx}

\newcommand{\boldface}[1]{\boldsymbol{#1}}  

\newcommand{\bfx}{\boldface{x}}
\newcommand{\bff}{\boldface{f}}
\newcommand{\bfp}{\boldface{p}}
\newcommand{\bfq}{\boldface{q}}

\newcommand{\bfz}{\boldface{z}}

\newcommand{\bfF}{\boldsymbol{F}}
\newcommand{\bfI}{\boldsymbol{I}}
\newcommand{\bfSigma}{\boldsymbol{\Sigma}}

\newcommand{\calZ}{\mathcal{Z}}
\newcommand{\T}{^{\mathrm{T}}} 

\def\dd{\;\!\mathrm{d}}

\DeclareSIUnit\angstrom{\text{Å}}

\begin{document}

\shorttitle{Predicting temperature-dependent failure and transformation zones in 2D silica}
\shortauthors{Sp\'inola et~al.}

\title[mode=title]{Predicting temperature-dependent failure and transformation zones in 2D silica glass through quasistatic Gaussian Phase Packets}
\author[eth]{Miguel Sp\'inola}[orcid=0000-0002-5180-6149]
\author[eth]{Shashank Saxena}[orcid=0000-0002-5242-9103]
\author[rwth]{Franz Bamer}
\cormark[1]
\ead{bamer@iam.rwth-aachen.de}
\author[eth]{Dennis M. Kochmann}[orcid=0000-0002-9112-6615]
\cormark[1]
\ead{dmk@ethz.ch}
\cortext[1]{Corresponding authors}
\address[eth]{Mechanics \& Materials Laboratory, Department of Mechanical and Process Engineering, ETH Z\"urich, Switzerland}
\address[rwth]{Institute of General Mechanics, RWTH-Aachen University, Germany}

\begin{abstract}
The athermal quasistatic (AQS) method is a powerful technique to study the mechanical behavior of disordered systems. However, its applicability is limited to temperatures near zero, where thermal activation is unlikely. In this work, we extend the AQS method to finite temperatures, based on a formulation that describes atoms as temperature-dependent Gaussian packets (GPPs) in phase space under quasistatic conditions, thus equivalent to minimum free energy conditions. This framework is used to study the effect of temperature on the onset of inelasticity and fracture in amorphous two-dimensional silica glass approaching quasistatic conditions under uniaxial tensile loading. An important characteristic of this formulation is the directional dependence of the variance of each Gaussian packet in configuration space, making this formulation an inexpensive and accurate predictor of zones prone to atomic-scale rearrangements, both in the undeformed state and (with increasing accuracy) as the deformation progresses. This method is also shown to accurately capture the thermal expansion of the disordered material. Furthermore, combining the GPP description with Metropolis sampling predicts the effect of temperature on the onset of fracture of the material, which is validated through MD simulations at strain rates as low as $10^{4}$s$^{-1}$. The presented framework therefore provides a valuable technique for studying the nonlinear mechanics of disordered materials at finite temperature and for predicting local rearrangement zones in disordered solids efficiently without the need for expensive MD simulations.
\end{abstract}


\maketitle

\section{Introduction}
 
Crystalline defects such as dislocations, stacking faults, or grain boundaries can be directly associated with different mechanisms promoting inelastic behavior in crystalline materials. For example, they result in plastic deformation of metallic materials, accommodating large deformation before failure. Disordered solids, by contrast, typically do not contain such localized and well-characterized defects. They do still exhibit plastic deformation, yet generally to a lesser degree and more commonly involving localization in thin shear bands, which are believed to result from shear transformation (ST) events --- structural rearrangements of clusters of atoms \cite{maloney_universal_2004}. Regions in which such rearrangements occur are also known as shear transformation zones (STZs) \cite{falk_dynamics_1998, patinet_connecting_2016}. STZs involve a few tens of atoms, and ST events span characteristic times of picoseconds \cite{rodney_modeling_2011}. They are accompanied by local heating, which results in decohesion of the material and catastrophic failure \cite{lewandowski_temperature_2006}. A better understanding of those processes can greatly benefit the development of glasses as engineering materials.

Unfortunately, the experimental study of such atomic rearrangements and, more generally, of the atomic structure of disordered materials has proven difficult due to the lack of long-range order in those materials. The structural characterization of such materials has mainly relied on X-ray diffraction, which only infers average atom distributions, making it difficult, if not impossible, to determine precise atomic structures \cite{roy2018silica, mozgawa_spectroscopic_2019}. About a decade ago, a two-dimensional (2D) version of silica (SiO$_2$) was characterized at atomic resolution, using numerical methods in conjunction with high-resolution transmission electron microscopy (TEM) \cite{huang_direct_2012,Lichtenstein2012,bjorkman_defects_2013}. Though this experimental system is a simplification of 3D reality, it can significantly aid the development and testing of atomistic models for the mechanical properties of glasses by serving as a benchmark platform. Following \citeauthor{ebrahem_vitreous_2020}, we will consider a numerical model that is topologically equivalent to experimentally observed 2D silica systems imaged by \citeauthor{Lichtenstein2012} \cite{Lichtenstein2012} to study the complete, nonlinear stress-strain response of a disordered network system by subjecting it to uniaxial deformation tests at finite temperature.

Numerical investigations of the stress-strain response of glasses at the atomic level have commonly used the Athermal Quasi-static Simulation (AQS) \cite{maloney_universal_2004, tanguy_plastic_2006, lerner_locality_2009, tsamados_local_2009, karmakar_statistical_2010, patinet_connecting_2016, ebrahem_vitreous_2020, ruscher_avalanches_2021, bamer2023molecular} or classical Molecular Dynamics (MD) simulations \cite{ochoa_molecular_1991, chowdhury_molecular_2016, wang_anomalous_2024, shekh_alshabab_criticality_2024}. The goal of the AQS is to simulate a quasi-static process with infinitely slow strain rates, thus overcoming one intrinsic issue when dealing with molecular dynamics -- the limited simulation time period. However, it neglects the effect of temperature, critical for triggering transitions in the potential energy surface (PES), by using Molecular Statics, making it a valid approximation at temperatures significantly lower than the glass transition temperature. By contrast, MD calculations can capture the effect of thermal fluctuations on transition events involving sudden rearrangements -- or STs in the particular case of shear deformation -- at higher temperatures, yet they are limited to high strain rates for practical computational times. Both strain rate and temperature play important roles in the mechanics of glasses \cite{ochoa_molecular_1991, pedone_molecular_2008, chowdhury_molecular_2016, wang_anomalous_2024}. \citeauthor{chowdhury_molecular_2016} summarized several MD studies on the mechanical properties of vitreous silica, whose lowest strain rate was $10^8~s^{-1}$. Another, more recent study investigated the fracture of metallic glasses at strain rates as low as $10^{7}~s^{-1}$ \cite{fan_effect_2024}. These rates are still far from those in most experimental studies. Therefore, there is a need for a numerical method that can bring down this gap by capturing the effects of temperature at lower strain rates on the onset of STs or atomic-scale rearrangements in general.

Predicting the location of such localized structural rearrangements is of high interest for understanding the effective material behavior. Many studies have used structural factors as indicators, such as those reviewed in \cite{richard_predicting_2020}. Another promising physically-based approach uses the vibrational modes to predict 'soft spots' \cite{widmer_cooper_localized_2009, ghosh_connecting_2011, manning_vibrational_2011, mosayebi_soft_2014, wu_topology_2023}. Those require the calculation of eigenvalues of the system's dynamical matrix, which comes with high computational costs when targeting larger systems.

In this work, we use a framework based on Gaussian Phase Packets (GPP), which resorts to statistical mechanics to tackle both of the aforementioned problems at once, while having a computational complexity close to $0K$ Molecular Statics. This framework assumes an ensemble probability distribution function (PDF) of Gaussian-type. The latter is parametrized by (i)~a covariance matrix of atomic positions related to the dynamical matrix of the system and (ii)~the set of atomic mean positions. The evolution of the PDF, taken to the quasi-static limit, yields a set of equations to be solved for the covariance matrix and mean positions in the free-energy-minimum state of the system, which is dependent on temperature, along, in principle, arbitrary deformation paths. Thus, it extends the zero-temperature AQS by including thermal expansion effects, as shown later. As a result, this procedure yields quasi-static states and individual atomic frequencies at once as a function of temperature. As this quasi-static setting alone cannot capture thermal fluctuations (such as those triggering STs or atomic-scale rearrangements in general), the method is further extended by a Metropolis algorithm with proposal probabilities informed by the Gaussian PDF.

The remainder of this contribution is organized as follows. Section~\ref{sec:Methods} introduces the AQS method and the GPP equations in the quasi-static limit with two variations: the hyperspherical (or \textit{isotropic}) Gaussian approximation and the non-hyperspherical (or \textit{anisotropic}) Gaussian approximation. Next, the Metropolis algorithm based on the equilibrium GPP probability distribution is described. In Section~\ref{sec:results}, two subsections present benchmark simulations using the new approach. First, the results of the GPP framework for the thermal expansion of silica samples are compared to MD data. These thermally expanded states serve as initial conditions for subsequent simulations. Second, GPP results from subjecting silica samples to uniaxial deformation are presented, comparing them to MD data across different strain rates. Uniaxial extension simulation results with and without the Metropolis algorithm are compared to the MD results, along with an analysis of tension-induced atomic-scale rearrangements, which can be seen as equivalents to shear-induced STs. Finally, Section~\ref{sec:conclusions} concludes this study, summarizes the main findings, and discusses the limits, advantages, and potential future improvements of the GPP framework for studying glasses and disordered solids in general.

\section{Methodology}
\label{sec:Methods}

We begin by introducing the Athermal Quasi-static Simulation (AQS) method and its extension to finite temperature via the quasi-static limit of the GPP framework, which models the system as an ensemble in the statistical mechanics sense by positing an ansatz for the ensemble probability distribution $f$. We use the latter to study the effect of temperature on the mechanical properties of 2D silica glass, a disordered benchmark problem having a complex disordered network structure, including the thermal expansion and the onset of fracture. First, we use it as a stand-alone framework and later combine it with a Metropolis algorithm with a trial function obtained from $f$. The Metropolis algorithm, to which we will refer as "Metropolis-GPP", is also presented at the end of this section. 

\subsection{The Athermal Quasi-Static (AQS) method}

The AQS method (see \cite{bamer2023molecular} for a detailed review) is a technique for modeling low-strain-rate conditions without the effect of thermal fluctuations. It is commonly chosen for modeling disordered solids, in particular glasses, at temperatures much lower than the glass transition temperature $T_g$. It consists of applying finite, small, affine deformation increments and is followed by minimization in the potential energy landscape (PEL) (or structural relaxations) in an alternating fashion. This generates a sequence of potential energy minima as the deformation progresses, mimicking a quasi-static process. Due to their lack of symmetry, disordered solids generally experience non-affine displacements during such relaxations, even in the elastic regime before atomic-scale rearrangements occur, contrary to crystalline materials. The atomic forces corresponding to those non-affine atomic displacements can be obtained from the Hessian of the potential energy $U$ with respect to the atomic positions (also known as the force-constants matrix) and the velocity field corresponding to the non-affine displacements (obtained by mapping the relaxed positions to the reference configuration). This is described in the following.

Consider a periodic system of $N$ atoms. We formulate molecular statics in a (total) Lagrangian framework. This way, one defines a reference domain $\hat{\mathcal{D}}$ with a reference simulation cell characterized by a Bravais tensor $\hat{H}$ and the atomic reference positions in the initial configuration, $\hat\bfq^{(0)}=\{\hat\bfq_{i}^{(0)}, \; i=1,\hdots,N\}\in\mathbb{R}^{dN}$ with $d$ referring to the dimension of the system. Furthermore, one defines a current domain $\mathcal{D}$ with the current simulation cell $\boldsymbol{H}$ and the atomic current positions $\bfq=\{\bfq_{i}, \; i=1,\hdots,N\}\in\mathbb{R}^{dN}$. Notably, $\hat\bfq^{(0)}$ and $\bfq$ are both constructed as concatenated vectors.
Mechanical loading is prescribed by introducing a mapping from $\hat{\boldsymbol{H}}$ to $\boldsymbol{H}$, written as:
\begin{equation}
\bfF = \hat{\boldsymbol{H}}^{-1}\boldsymbol{H}.
\end{equation}
The tensor $\bfF$ performing this mapping may also be seen as an equivalent to the deformation gradient in continuum mechanics. Equivalent to the positions, we define the global representation of $\boldsymbol{F}$ as
\begin{equation}
    \bar{\boldsymbol{F}} =
    \left(
    \begin{array}{ccc}
        \boldsymbol{F} &        &                  \\
                       & \ddots &                  \\
                       &        & \boldsymbol{F}
    \end{array}
    \right) \in \mathbb{R}^{dN\times dN},
\end{equation}
allowing for affine mappings of concatenated position vectors.
Any particle position $\bfq_i$ in a deformed cell $\boldsymbol{H}$ can be transformed back into its corresponding reference position $\hat{\bfq}_i$ located in $\hat{\boldsymbol{H}}$. We note that such a pull-back operation does, in general, not lead to the initial undeformed configuration, i.e., $\hat{\boldsymbol{q}}_i^{(0)} \neq \boldsymbol{F}^{-1}\boldsymbol{q}_i$. This would only be the case for affine deformation as found, e.g., at small strains in centro-symmetric ordered structures.
Since we are observing our system from a Lagrangian perspective, we describe mechanical quantities of interest with respect to the reference domain.

Initially, the system is located in the reference domain with the initial atomic reference positions $\hat\bfq^{(0)}$ in the reference cell $\hat{\boldsymbol{H}}$. $\hat{\boldsymbol{H}}=\left[\hat{\boldsymbol{l}}_1,\hat{\boldsymbol{l}}_2\right]$ must be invertible, and it consists of the linearly independent Bravais vectors $\boldsymbol{l}_1$ and $\boldsymbol{l}_2$ in 2D.
At zero temperature, the system rests in a minimum of the PES, so the force $\hat{\bff}_i$ on every particle $i$ vanishes:
\begin{equation}
    \hat{\bff}_i = -\nabla_{\hat{\bfq}_i} \hat{U}\left(\hat{\bfq},\boldsymbol{H}\right) =-\bfF^\top\nabla_{\bfq_i}U(\bfq,\boldsymbol{H})= \boldsymbol{0},
\end{equation}
where we also measure the potential energy with respect to the reference configuration, i.e., $\hat{U}\left(\hat{\bfq},\boldsymbol{H}\right) = U\left(\boldsymbol{F}\hat{\bfq},\boldsymbol{H}\right)$.

Deformation is induced by altering the shape of the deformation cell $\hat{\boldsymbol{H}}$ to $\boldsymbol{H}$ and displacing the atomic positions located in the cell affinely. After this external disturbance, we require that the configuration remains in a minimum of the potential energy surface, restricting any inertia and introducing overdamped behavior. Consequently, the forces $\hat\bff=\left\{\hat\bff_i, i=1,\ldots,N\right\}$ on all atoms remain zero during deformation, i.e.,
\begin{equation}
\hat\bff(\boldsymbol{H}) = -\nabla_{\hat\bfq} \hat{U}(\hat\bfq,\boldsymbol{H})= -\bfF^\top\nabla_{\bfq}U(\bfq,\boldsymbol{H})=\boldsymbol{0}.
\end{equation}
This force equilibrium is extended by requiring that the change in force with deformation is also zero:
\begin{equation}
    \frac{d \hat{\bff}}{d\boldsymbol{H}} = \frac{\partial \hat{\bff}}{\partial \boldsymbol{H}} + \hat{\boldsymbol{\mathcal{H}}}\left(\hat{\bfq},\boldsymbol{H}\right)\frac{\partial\hat{\boldsymbol{q}}}{\partial \boldsymbol{H}} = \boldsymbol{0} ,
\end{equation}
where we consider that $U$ depends on both $\hat{\bfq}$ and $\boldsymbol{H}$. In this equation, $\hat{\boldsymbol{\mathcal{H}}}(\hat{\bfq},\boldsymbol{H})=\frac{\partial \hat{U}(\hat\bfq,\boldsymbol{H})}{\partial \hat\bfq\partial \hat\bfq} \in \mathbb{R}^{dN\times dN}$ refers to the Hessian of the potential energy landscape. We furthermore interpret the term $\Tilde{\bff} = \frac{\partial \hat{\bff}}{\partial \boldsymbol{H}}$ as a virtual force that pushes the system back into the local minimum of the PES.
Thus, the equilibrium equation for molecular systems is written as
\begin{equation}
\tilde\bff = \hat{\boldsymbol{\mathcal{H}}}\left(\hat{\bfq},\boldsymbol{H}\right)\frac{\partial\hat\bfq}{\partial\boldsymbol{H}},
\label{eq:nonaff_correc}
\end{equation}
where $\frac{\partial\hat\bfq}{\partial\boldsymbol{H}}$ is the change in the atomic positions pulled back to the reference configuration, i.e., the change in atomic positions without the affine disturbance. Thus, it is generally a non-affine displacement field \cite{maloney_universal_2004, maloney_amorphous_2006, bamer2023molecular}. Equation \eqref{eq:nonaff_correc} reveals that, when disturbing a system by external mechanical deformation, a virtual force of $\Tilde{\bff} = \frac{\partial \hat{\bff}}{\partial \boldsymbol{H}}$ is necessary to push the system back into the minimum along the non-affine direction $\frac{\partial\hat\bfq}{\partial\boldsymbol{H}}$.

An AQS protocol is implemented as a sequence of deformation increments, each consisting of two alternating steps. In step one, $\bfF$ is imposed by affinely displacing all atoms and altering the Bravais vectors of the simulation cell. In step two, the atomic positions are relaxed to their equilibrium positions, while holding the deformed simulation box fixed. This is mainly realized by appropriate minimization algorithms. Consequently, the deformed relaxed atomic positions and the corresponding deformed simulation box are written as, respectively,
\begin{align}\label{eq:rCurrent}
    \bfq_i &=\bfF \hat\bfq_{i}^{(0)}+\bfx_i,\\
    \boldsymbol{H} &= \boldsymbol{F}\hat{\boldsymbol{H}},
\end{align}
where vector $\bfx_i$ is a finite non-affine atomic displacement correction of atom $i$. Mapping the current deformed position \eqref{eq:rCurrent} to its representation in the reference configuration results in
\begin{equation}
    \hat\bfq_i=\bfF^{-1}\bfq_i=\hat\bfq_{i}^{(0)}+\hat\bfx_i.
\end{equation}
We again note that the structural relaxations in the AQS method assume zero-temperature conditions, making it only applicable to problems at lower temperatures. In this work, we propose an extension to finite temperature, as discussed in the following.

\subsection{The framework of Gaussian Phase Packets}

The Gaussian Phase Packets (GPP) framework is a suitable candidate for extending the AQS method to finite temperature, as it operates with statistical ensemble parameters of atomic degrees of freedom instead of instantaneous ones and can be taken to the quasi-static limit. We will utilize the GPP framework to solve for the equilibrium ensemble probability density at constant temperature, which is assumed to be of Gaussian-type parametrized by the atomic mean positions, mean momenta, and their covariance matrix. The AQS protocol is therefore extended to finite temperature by solving for the finite temperature equilibrium ensembles (approximated by the GPP framework) visited during each deformation increment to realize quasi-static conditions. In the following, we introduce the GPP framework in its quasi-static limit. Detailed derivations and explanations of the GPP formulation (in its simpler, isotropic form and without relation to the AQS method) can be found in \citeauthor{gpp2021}~\cite{gpp2021}. 

For conciseness, we use the condensed notation $\bfz=\left(\bfp(t),\bfq(t)\right)  \in \mathbb{R}^{2dN}$ to refer to the phase-space coordinates of a system of $N$ atoms in $d$ dimensions. As implied by the name, the GPP framework assumes that our ensemble is characterized at time $t$ by a probability density $f$ of Gaussian type, specifically defining
\begin{equation}\label{eq_gpp_full}
     f(\bfz ,t) = \frac{1}{\calZ(t)} \exp\left[ -\frac{1}{2}\left(\bfz  - \bar{\bfz }(t)\right)\T \boldsymbol{\Sigma}^{-1}(t) \left(\bfz  - \bar{\bfz }(t)\right)\right].
\end{equation}
This distribution provides the probability of finding our system in a volume $\dd\bfz$ at time $t$, where $\calZ(t)$ is the partition function defined via $\int f(\bfz,t)\dd \bfz = \langle 1\rangle = 1$ by integrating over all of phase space $\Gamma$ (denoting phase-space averages by $\langle\cdot\rangle=\int_\Gamma(\cdot)f(\bfz,t)\dd\bfz$). This probability density is parametrized by the mean phase-space coordinates
\begin{equation}
    \bar{\bfz }(t) = \langle\bfz\rangle=\int_\Gamma \bfz f(\bfz,t)\dd\bfz
\end{equation}
and the covariance matrix of interatomic positions and momenta, denoted by $\bfSigma\in \mathbb{R}^{2dN\times 2dN}$. This matrix is symmetric and composed of three unique submatrices, written as
\begin{equation}
    \bfSigma = 
    \begin{bmatrix}
        \bfSigma^{(\bfp,\bfp)} & \bfSigma^{(\bfp,\bfq)} \\
        \bfSigma^{(\bfp,\bfq)} & \bfSigma^{(\bfq,\bfq)}
    \end{bmatrix},
\end{equation}
where each submatrix is of size $dN\times dN$ and contains the correlation coefficients between the different atomic degrees of freedom. Any thermodynamic observable of interest can be obtained as an integral of the corresponding observable weighted by the distribution function $f(\bfz,t)$ over the entire phase space. Hence, determining the statistical parameters of the probability distribution function is the main objective of this approach --- as opposed to computing the evolution of the instantaneous degrees of freedom of a single microstate as, e.g., in MD. The evolution equations for these statistical parameters can be obtained by inserting the ansatz~\eqref{eq_gpp_full} into Liouville's equation, which describes the dynamics of the probability density $f(\boldsymbol{z},t)$ of a Hamiltonian system \cite{gibbs1902elementary, romero_extended_2023}. Since solving for the evolution of all parameters in the covariance matrix $\boldsymbol{\Sigma}(t)$ becomes computationally intractable for large systems, we resort to an approximation of the probability distribution by assuming interatomic independence, which transforms the total probability distribution function into a product of individual (atomic) Gaussian phase packets, according to 
\begin{align}\label{eq_gpp_indep}
     f(\bfz ,t) = \frac{1}{M}\prod_{i=1}^N f_i(\bfz _i,t)
     \qquad\text{with}\qquad
     f_i(\bfz _i,t) = \frac{1}{\calZ _i(t)} \exp\left[ -\frac{1}{2}\left(\bfz _i - \bar{\bfz _i}(t)\right)^\top \boldsymbol{\Sigma}_i^{-1}(t) \left(\bfz _i - \bar{\bfz _i}(t)\right)\right],
\end{align}
where $M$ is a normalization constant (stemming from the partition function $\mathcal{Z}(t)$), which becomes $M=N!$ for a system with a single atom type and $M=N_1!N_2!$ for a system with two types of atoms -- $N_1$ atoms of one type and $N_2$ of the other.

As $\bfSigma_i$ is symmetric, it is composed of three unique submatrices. In this work, we study a 2D system, resulting in the following submatrices defined with respect to the Cartesian coordinates $x$ and $y$:
\begin{equation}
\label{eq_submatrices}
\boldsymbol{\Sigma}^{(\bfp,\bfp)}_{i} = 
\begin{bmatrix}
\Omega_i^{xx} & \Omega_i^{xy} \\
\Omega_i^{xy} & \Omega_i^{yy} \\
\end{bmatrix}, \qquad
\boldsymbol{\Sigma}^{(\bfq,\bfq)}_{i}=
\begin{bmatrix}
\Sigma_i^{xx} & \Sigma_i^{xy} \\
\Sigma_i^{xy} & \Sigma_i^{yy}\\
\end{bmatrix},\qquad\text{and}\quad
\boldsymbol{\Sigma}^{(\bfp,\bfq)}_{i}= \boldsymbol{\Sigma}^{(\bfq,\bfp)}_{i}
\begin{bmatrix}
\beta_i^{xx} & \beta_i^{xy} \\
\beta_i^{xy} & \beta_i^{yy}
\end{bmatrix}
\end{equation}
with coefficients
\begin{equation}
\begin{split}
    \Omega_i^{\alpha\beta} &= \langle (\bfp_i^\alpha - \bar{\bfp _i}^\alpha) \otimes (\bfp_i^\beta - \bar{\bfp _i}^\beta) \rangle, \\ 
    \Sigma_i^{\alpha\beta} &= \langle (\bfq_i^\alpha - \bar{\bfq _i}^\alpha) \otimes (\bfq_i^\beta - \bar{\bfq _i}^\beta) \rangle, \\ 
    \beta_i^{\alpha\beta} &= \langle (\bfp_i^\alpha - \bar{\bfp _i}^\alpha) \otimes (\bfq_i^\beta - \bar{\bfq _i}^\beta) \rangle.
\end{split}
\end{equation}

Previous works leveraging the GPP \cite{gpp2021, saxena2022fast, spinola2024finite} and related formulations \cite{kulkarni2008variational, li2011diffusive, lesar1989finite} made a further approximation and assumed $f_i$ to be of hyperspherical shape (which we refer to as an \textit{isotropic} Gaussian). This yields diagonal correlation submatrices
\begin{equation}
    \bfSigma^{(\bfp,\bfp)}_i = \Omega_i \bfI, \quad
    \bfSigma^{(\bfq,\bfq)}_i = \Sigma_i \bfI, \quad
    \bfSigma^{(\bfq,\bfp)}_i = \beta_i\bfI
\end{equation}
with scalars $\Omega_i$, $\Sigma_i$, and $\beta_i$ (characterizing the atomic self-correlation of atom $i$). Under this assumption, the set of parameters of every atomic site is reduced to $\{\bar{\bfp}_i, \bar{\bfq}_i,  \Omega_i, \Sigma_i, \beta_i  \}$. In principle, these parameters are time-dependent. To arrive at the thermodynamic equilibrium configuration of the atomic ensemble, which corresponds to a time-independent probability density $f$, we take the quasi-static limit of the evolution equations for the above statistical parameters. In this limit, the mean momentum $\bar{\bfp}_i$ and mean thermal momentum $\beta_i$ vanish for every atomic site~$i$ \cite{gpp2021}. The resulting equilibrium equations for every atom $i$ under the interatomic-independent and isotropic Gaussian distribution are
\begin{subequations}\label{eq_gpp_iso}
    \begin{align}
        &\langle \bff_i \rangle = \bf{0}\qquad \mathrm{and} \label{gpp_iso_a}\\
        &\frac{\Omega_i}{m_i} + \frac{ \langle \bff_i(\bfq ) \cdot (\bfq_i  - \bar{\bfq}_i)  \rangle }{3} = 0, \label{gpp_iso_b}
    \end{align}
\end{subequations}
where $\bff_i$ denotes the net mechanical force on atom $i$ and $m_i$ is the mass of atom $i$. These equations correspond to the vanishing of the mean mechanical forces (Eq.~\eqref{gpp_iso_a}) and mean thermal forces (Eq.~\eqref{gpp_iso_b}). The momentum correlation coefficients ($\Omega_i^{\alpha\beta}$) are obtained by imposing the thermodynamic process under which the system is equilibrated. In this work, we are interested in isothermal conditions specified by a constant (absolute) temperature $T$. This results in $\Omega_i=m_ik_BT$ (see \cite{gpp2021}), where $k_B$ is Boltzmann's constant. Under this condition, the GPP probability density in 2D assumes the form
\begin{equation}
\begin{aligned}
f_i(\bfq_i,\bfp_i;\overline\bfq_i,\bfSigma_i^{(\bfq,\bfq)},T) 
&= \underbrace{\frac{1}{\mathcal{Z}_{i}^{\bfq}}
\text{exp}
\left(
-\frac{1}{2}(\bfq_i-\overline\bfq_i)^\top\left[\bfSigma_i^{(\bfq,\bfq)}\right]^{-1}(\bfq_i-\overline\bfq_i)
\right)}_{=f^{\bfq}_i(\bfq_i;\overline\bfq_i,\bfSigma_i^{(\bfq,\bfq)})}
\underbrace{
\frac{1}{\mathcal{Z}^{\bfp}_i}
\text{exp}
\left(
-\frac{1}{2}\frac{|\bfp_i|^2}{m_ik_BT}
\right)}_{=f_i^{\bfp}(\bfp_i;T)}
\end{aligned}
\label{eq:full_pdf_i_iso}
\end{equation}
with
\begin{equation}
    \mathcal{Z}_i=\mathcal{Z}_{i}^{\bfq}\mathcal{Z}_i^{\bfp}=
    \left(\frac{2\pi}{h}\right)^2 m_ik_BT\sqrt{\det\bfSigma^{(\bfq,\bfq)}_i}
\end{equation}
and $h$ denoting Planck's constant. Thus, under isothermal conditions with a given temperature $T$, Eqs.~\eqref{eq_gpp_iso} with $\Omega_i=m_ik_BT$ (for $i=1,\ldots,N$) are to be solved for the atomic mean positions $\bar{\bfq}=\{\bar{\bfq}_{i} : i=1,\ldots,N\}$ and position variances $\Sigma=\{\Sigma_i : i=1,\ldots,N\}$ to fully determine $f_i$. Note that this solution can be reinterpreted \cite{gpp2021} as the minimizer of the free energy generated by the probability density in~\eqref{eq_gpp_indep}.

As prior work based on the GPP framework studied crystalline materials with long-range order \cite{gpp2021, saxena2022fast, spinola2024finite}, the isotropic Gaussian assumption was shown to be appropriate in those specific cases. This, however, is no longer the case when studying amorphous systems, as we will show later in this paper. Therefore, in this work, the isotropic Gaussian assumption is relaxed to increase simulation accuracy, which is critical for predicting the location of inelastic events in the silica samples studied below. This relaxed assumption in 2D assumes the position variance (sub-)matrix to be 
\begin{equation}
    \bfSigma^{(\bfq,\bfq)}_i = \begin{bmatrix}
        \Sigma_i^{xx} & \Sigma_i^{xy}\\
        \Sigma_i^{xy} & \Sigma_i^{yy}
    \end{bmatrix},
\end{equation}
now described by the three independent scalars $\Sigma_i^{xx}$, $\Sigma_i^{yy}$, and $\Sigma_i^{xy}$ (instead of a single scalar as in the isotropic case). Consequently, the equilibrium equations for atom $i$ (with $\alpha,\beta\in\{x,y\}$ and $\bf{v}^\alpha$ denoting the $\alpha$th component of vector $\bf{v}$) become
\begin{subequations}\label{eq_gpp_aniso}
     \begin{align}
        &\langle \bff_i \rangle = \bf{0}\qquad \mathrm{and}\\
        &\frac{\Omega^{\alpha \beta}_i}{m_i}\delta_{\alpha\beta} + \frac{\langle f^\alpha_i(q^\beta_i - \bar{q}^\beta_i\rangle + f^\beta_i(q^\alpha_i - \bar{q}^\alpha_i\rangle)}{2}=0,
    \end{align}
\end{subequations}
with $\delta_{\alpha\beta}$ referring to the Kronecker delta.
Imposing isothermal conditions in this case requires $\Omega^{\alpha\beta}=m_ik_BT\delta_{\alpha\beta}$, according to the equipartition theorem \cite{tadmor2011modeling}. The solution to Eq.~\eqref{eq_gpp_aniso} under the aforementioned isothermal conditions is a set of mean positions, $\bar{\bfq}=\{\bar{\bfq}_{i} : i=1,\ldots,N\}$, and a set of (anisotorpic) position variances, $\bfSigma^{(\bfq,\bfq)}=\{\bfSigma_{i}^{(\bfq,\bfq)} : i=1,\ldots,N\}$. 

In practice, Eqs.~\eqref{eq_gpp_aniso} and \eqref{eq_gpp_iso} (for the isotropic and anisotropic case, respectively) are solved numerically using the Fast Inertial Relaxation Engine (FIRE) \cite{fire}, starting from an initial guess for $\{\overline\bfq,\bfSigma^{(\bfq,\bfq)}\}$. Phase averages $\langle \cdot \rangle$ are computed using a third-order Gaussian quadrature rule in phase space \citep{stroud_approximate_1971,saxena2023gnn}. For the anisotropic Gaussian approximation, the quadrature points for each Gaussian phase packet are taken in the principal directions of $\bfSigma^{\bfq\bfq}_i$, which significantly improves the quadrature accuracy (compared to keeping the directions fixed). Furthermore, to ensure that $\bfSigma^{\bfq\bfq}_i$ always remains positive-definite during FIRE relaxation, the eigenvalues of $\bfSigma^{\bfq\bfq}_i$ are monitored and, if they become negative, the timestep is reduced in an iterative fashion, until the eigenvalues become positive again.

To determine the stress corresponding to a relaxed, deformed configuration, we compute the virial stress tensor.
For an ensemble defined by Eq.~\eqref{eq_gpp_indep}, the stress tensor $\boldsymbol{\sigma}$ is obtained from the derivative of the free energy of the ensemble with respect to deformation gradient $\boldsymbol{F}$ (mapping a reference state of the system to a deformed state, both dependent on the mean positions). This results in
\begin{equation}\label{eq_vir_gpp}
\sigma_{\alpha\beta} = \frac{1}{V}\sum_{i=1}^N\sum_{
j\neq i
} \left\langle f^{\alpha}_{ij} \left(\bar{q}^{\beta}_j - \bar{q}^{\beta}_i\right)\right\rangle = \frac{1}{V}\sum_{i=1}^N\sum_{
j\neq i
} \langle f^{\alpha}_{ij} \rangle\left(\bar{q}^{\beta}_j - \bar{q}^{\beta}_i\right),
\end{equation}
where $V$ is the volume of the simulation box and $f^{\alpha}_{ij}$ is the $\alpha$-component of the force felt by atom $i$ due to atom $j$, separated by a distance $r_{ij}$. For more information about this derivation, we refer to \citeauthor{tadmor2011modeling} \cite{tadmor2011modeling}. The $\alpha$-component of the force is obtained from the interatomic potential $U$ as:
\begin{equation}
\langle f_{ij}^{\alpha} \rangle = \left\langle
\frac{\partial U_i}{\partial \bf r_j}
 \right\rangle.
\end{equation}
Evaluating Eq.~\eqref{eq_vir_gpp} for a deformed simulation box admits constructing stress-strain curves, as demonstrated in our benchmark examples.

Let us highlight the differences between the MD equations of motion and the GPP quasi-static equations. Within the GPP framework, for a given temperature $T$ the parameters $\{\overline\bfq,\bfSigma^{(\bfq,\bfq)}\}$ define a point in the free energy landscape $\mathcal{F}_T(\overline\bfq,\bfSigma^{(\bfq,\bfq)})$, and this point (or macrostate) characterizes a specific ensemble of microstates. Each microstate corresponds to a set of atomic positions denoted by $\bfq=\{\bfq_1,\hdots,\bfq_N\}$ in configuration space, and it has a probability $f^{\bfq}(\bfq;\overline\bfq,\bfSigma^{(\bfq,\bfq)})$. The negative gradient of the potential energy surface (PES) in configuration space, evaluated at a particular microstate $\bfq$, yields generalized instantaneous forces $\bff(\bfq)=\{\bff_1, \ldots,\bff_N\}$ composed of the (net) forces on all $N$ atoms within that microstate. MD tracks the atomic system over time as it visits various microstates, driven by $\bff(\bfq)$. In the GPP framework, instead, the generalized forces of all microstates corresponding to a particular macrostate in the PES are calculated and weighted by their probability $f^{\bfq}(\bfq;\overline\bfq,\bfSigma^{(\bfq,\bfq)})$ to obtain generalized \textit{phase-averaged} forces $\langle \bff\rangle= \{\langle \bff_0 \rangle, \ldots, \langle \bff_N\rangle\}$. Individual microstates are not evolved; instead, those phase-averaged forces are used to evolve $f^{\bfq}(\bfq; \overline\bfq,\bfSigma^{(\bfq,\bfq)})$ by translating $\overline\bfq$ in the free energy surface, until it becomes stationary (satisfying \eqref{eq_gpp_iso} or \eqref{eq_gpp_aniso}). Analogously, the thermal forces drive $\bfSigma^{(\bfq,\bfq)}$, so that the resulting free energy is minimized. This can be interpreted as evolving the Gaussian $f$, extending over many microstates in the PES, which evolves the probability assigned to each microstate, with maximum probability for the microstate with $\bfq=\overline\bfq$. In summary, MD traces a trajectory in configuration space, never settling in a stationary state at finite temperature, while the GPP model evolves an ensemble of microstates distributed according to a Gaussian probability density in configuration space by evolving $\{\overline\bfq,\bfSigma^{(\bfq,\bfq)}\}$ until arriving at a stationary point in the free energy landscape. Under the ergodic assumption assumed in the GPP framework, the time averages of the (equilibrated) MD trajectory are approximately equal to the phase-averages of the stationary probability density of the GPP.

\subsection{Sampling microstates about the equilibrium}
\label{sec:MC}

For regions of the PES with adjacent minima and associated transitional states, the dynamics simulated by MD can drive a microstate across a potential energy barrier into an adjacent PES minimum. Following harmonic transition state theory \cite{vineyard1957frequency}, the probability for a transition to occur is proportional to $e^{-\Delta U/k_BT}$, where $\Delta U$ is the height of the potential energy barrier between the transition state and the adjacent PES minimum. For the system studied in this work, such transitions occur, e.g., when an Si-O bond breaks. This probability is non-zero even for large energy barriers, so that for sufficiently long times (assuming ergodicity) the transition is observed in MD. Perturbations, such as affine deformations in this work, can increase this probability and the associated transition rate. 
In the GPP framework, by contrast, a system in thermodynamic equilibrium is associated with a static $f$, which is what we solve for in Eqs.~\eqref{eq_gpp_iso} and \eqref{eq_gpp_aniso}. While in MD a microstate under the canonical ensemble oscillating around a PES minimum will transition to nearby ones (given sufficient time), for the GPP-obtained $f$, which is necessarily centered near the PES minimum, which results in a free energy minimum per PES minimum, a (artificial) perturbation is necessary to overcome the resulting free energy barrier. As a result, whereas in MD perturbations to $\bfq$ are assisted by thermal fluctuations, thus allowing for transitions to adjacent PES minima, a transition of $f$ under the GPP framework between free energy minima must be triggered by perturbations to $\overline\bfq$.

To this end, a method is proposed for producing new initial conditions for the set of statistical parameters $\{\overline\bfq,\bfSigma\}$ close to a previous set but potentially sufficiently far, so that a statistical majority of the microstates of the resulting new initial conditions define an ensemble where a majority of microstates contributing to the generalized forces $\langle \bff \rangle$ drive the ensemble to an adjacent free energy minimum. While there are many options for such a method, in this work we produce a Markov chain, using the Metropolis algorithm \cite{metropolis1953equation} for trial microstates close to a given GPP ensemble by using the relaxed GPP probability distribution $f(\overline\bfq,\bfSigma)$ as a trial probability. This approach introduces a natural trial probability with potentially high acceptance rates. As we are interested in simulating the canonical ensemble, the acceptance probability will be a function of the probability density $f_\text{NVT}$ of the true canonical ensemble, as described below. This combination allows us to sample microstates close to the mean of a given ensemble characterized by $f(\overline\bfq,\bfSigma)$, including those at potential adjacent free energy minima, while being consistent with the true canonical ensemble of the system. The last microstate in this chain is chosen as the new mean of the GPP ansatz, while $\bfSigma$ is left unchanged. In combination, they are used as a new starting guess for minimizing the free energy and potentially capturing adjacent minimum free energy states.

The Markov chain is defined by the probability $R(x|y)$ of a system in microstate $y$ to transition to microstate $x$. This transition probability results from the proposal probability $T(x|y)$ and the acceptance probability $A(x|y) $ according to \cite{tuckerman2023statistical}
\begin{equation}
    R(x|y)=A(x|y)T(x|y),
\label{eq:transition_prob}
\end{equation}
where $T(x|y)$ is chosen to be the probability of microstate $x$ given a GPP ensemble with mean $\overline\bfq$ and variance $\bfSigma$. The transition probability must satisfy the detailed balance equation
\begin{equation}
    R(x|y)F(y) = R(y|x)F(x)
    \label{eq:detailed_balance}
\end{equation}
for unbiased sampling, where $F(x)$ is the probability of observing microstate $x$ for given thermodynamic conditions. To sample microstates consistent with the canonical ensemble, we choose $F(x)=f_\text{NVT}(x)$. Inserting \eqref{eq:transition_prob} into \eqref{eq:detailed_balance} results in the acceptance probability
\begin{equation}
    A(x|y) = \frac{T(y|x)f_\text{NVT}(x)}{T(x|y)f_\text{NVT}(y)}A(y|x) = r(x|y)A(y|x).
\end{equation}
For symmetric proposal probabilities like in the GPP ansatz, $T(x|y)=T(y|x)$ and thus the acceptance ratio is
\begin{equation}
    r(x|y) = \frac{f_\text{NVT}(x)}{f_\text{NVT}(y)} = \text{e}^{-\frac{U(x)-U(y)}{k_BT}},
\end{equation}
which yields the choice of the acceptance probability characteristic of the Metropolis algorithm \cite{metropolis1953equation} as:
\begin{equation}
    A(x|y) = \text{min}\left[1, \text{e}^{-\frac{U(x)-U(y)}{k_BT}}\right].
    \label{eq:accep_prob}
\end{equation}
In our numerical procedure, the generalized vectors are $x=\bfq_{k+1}$ and $y=\bfq_{k}$, where $k$ denotes the state $k$ in the Markov chain. Therefore,
\begin{equation}
    T(\bfq_{k+1}|\bfq_k) = f(\bfq_{k+1};\overline{\bfq}=\bfq_k,\bfSigma).
\end{equation}
This means that the GPP ansatz mean is shifted to the microstate selected after a trial, while the acceptance probability depends on the energies of the microstates $\bfq_{k+1}$ and $\bfq_k$. After $N_s$ trials, the starting guess for minimizing the free energy of the ensemble is $\{\overline\bfq=\bfq_{N_s},\bfSigma\}$. In our numerical examples, each of the $N_s$ trials displaces a single atom and calculates its acceptance probability according to Eq.~\eqref{eq:accep_prob}. These $N_s$ trials are performed in $N_p$ passes, each pass consisting of a trial move for every atom. Therefore, $N_s=N_pN$ where $N$ is the number of atoms in the system.

\section{Results and Discussion}
\label{sec:results}

To demonstrate the GPP framework introduced above, we aim to efficiently predict quasi-static finite-temperature effects on the mechanics of 2D disordered network glass structures that are topologically equivalent to imaged silica network structures from experiments \cite{buchner2014ultrathin,zhong2022two}. To this end, simulations that trigger the onset of inelastic behavior at different temperatures at increasingly low uniaxial strain rates were carried out for different network glass samples -- using the GPP framework and MD simulations for comparison. While in MD the strain rate is set by the integration timestep and the applied deformation increments, in the GPP framework in the quasi-static limit, presented in Eqs.~\eqref{eq_gpp_aniso} and \eqref{eq_gpp_iso}, there is no time dependence and hence no finite strain rate. However, the Metropolis sampling introduced above equips the GPP framework with an effective strain rate, as we will explain and illustrate in this section, where we describe the simulation setups and then the obtained results for, first, thermal expansion (comparing the isotropic and anisotropic GPP assumptions) and, second, numerical uniaxial loading experiments.

\begin{figure}[!htb]
\centering
\begin{tabular}{ccc}
\includegraphics[scale=0.4]{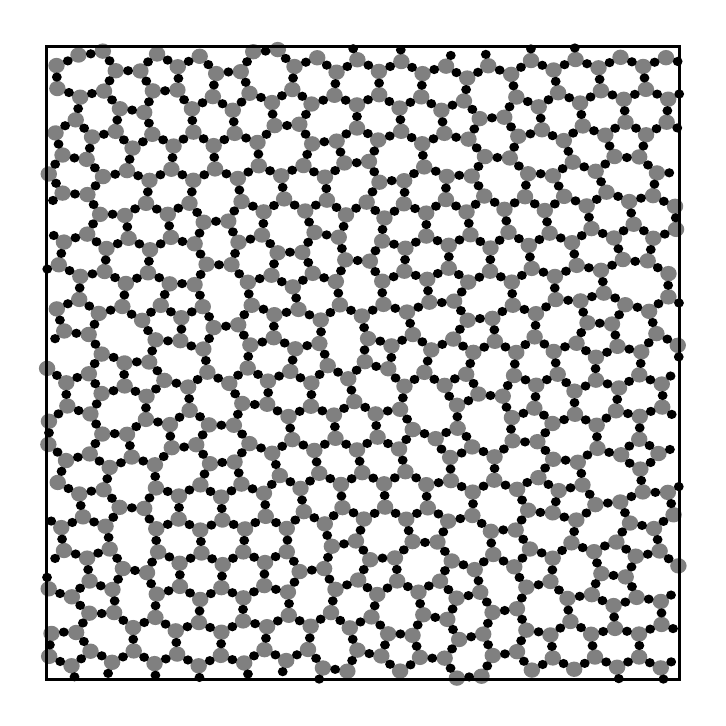}&
\includegraphics[scale=0.4]{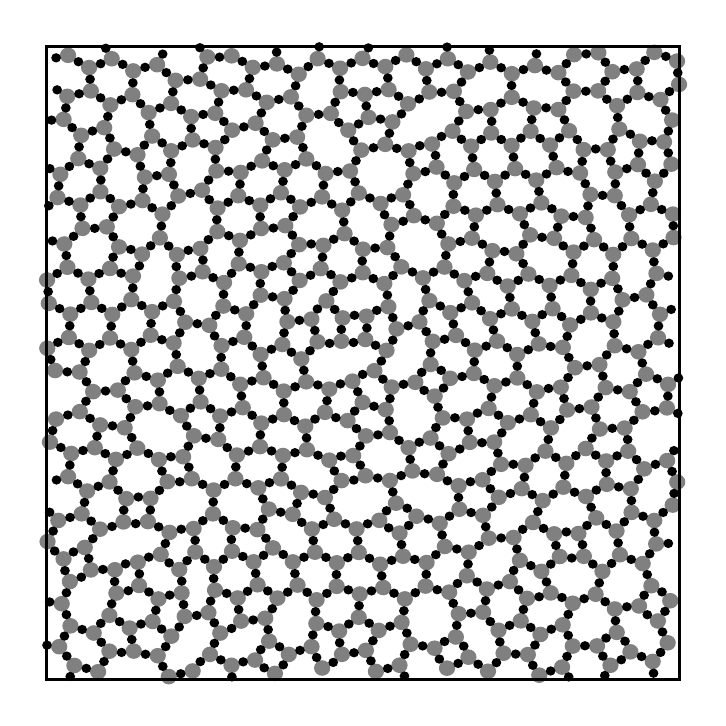}& 
\includegraphics[scale=0.4]{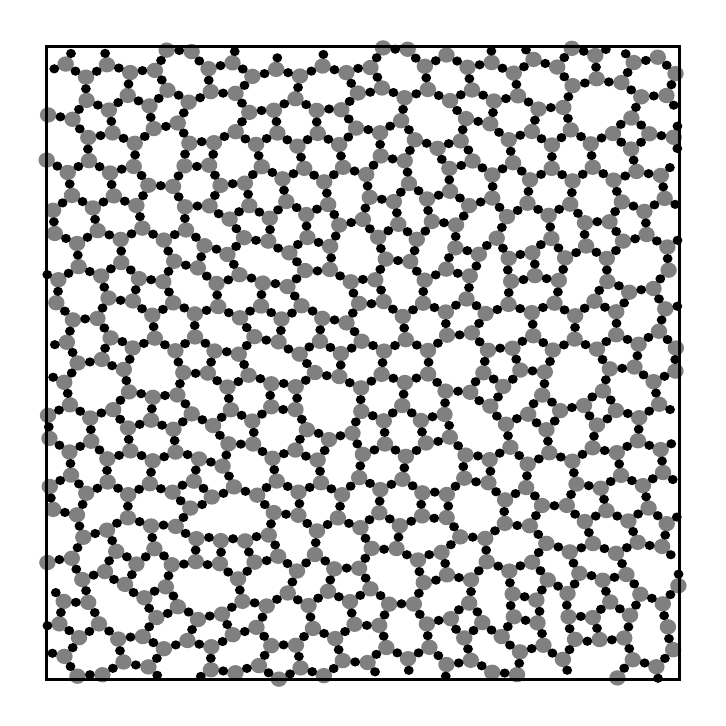} \\
(a) & (b) & (c)
\end{tabular}
\caption{2D silica samples, each containing 1350 atoms, used in this work, generated with the Monte-Carlo bond switching algorithm \cite{bamer2023molecular} with increasing number of bond-swtiching attempts (and hence increasing level of disorder) from left to right.}
\label{fig:samples_initial}
\end{figure}

Simulations were performed on samples generated using the Monte Carlo bond switching algorithm elaborately described in \cite{bamer2023molecular} with different numbers of topological transformation sequences resulting in various degrees of disorder. Starting from the hexagonal lattice structure, switching sequences are realized following a Monte Carlo Markov Chain approach. This way, every switch is accepted or rejected based on an objective function that quantifies the current network topology with the target network topology taken from experimental images \cite{Lichtenstein2012,huang_direct_2012}. Beyond the topological acceptance criterion, we included a purely physical acceptance criterion, which ensures that all SiO$_3$ triangles remain fully coordinated in the network, thereby maintaining the ring structure in accordance with the experimental images. Generating topologically equivalent samples numerically allows us to consider periodic boundary conditions in both dimensions with the simulation box defined by vectors $\boldsymbol{A}=[L_x,0],\boldsymbol{B}=[\gamma_0, L_y]$. Their dimensions range between $77.15\text{\AA}\times82.30\text{\AA}$ and $77.21\text{\AA}\times82.33\text{\AA}$ for $L_x$ and $L_y$, respectively, before relaxation, with a total of $1350$ atoms. Figure \ref{fig:samples_initial} shows three samples with increasing disorder.

The atomic interactions of the network glass system are modeled by a Yukawa-type pair-potential \cite{roy2018silica} written as:
\begin{equation}\label{eq_yukawa_V}
V(r,s_i,s_j) = \left(\frac{\sigma(s_i,s_j)}{r}\right)^{12} + \frac{q(s_i,s_j)}{r}\text{exp}(-\kappa/r),
\end{equation}
where $r$ is the distance between the pair of atoms $i$ and $j$ of species $s_i$ and $s_j$, respectively. The dimensionless parameters for silica are those used by \cite{roy2018silica}. They are $q(1,1)=1.5$, $q(1,2)=-1$, $q(2,2)=0.67$ to ensure charge neutrality, and the energy scale is set by the parameters $\kappa=1/5.649$, $\sigma(1,1)=2.25$, $\sigma(1,2)=1.075$, $\sigma(2,2)=0.9$, where $s_i=1$ refers to silicon (Si) and $s_i=2$ to oxygen (O). Following \citeauthor{roy2018silica}~\cite{roy2018silica} (additional data), the unit distance is $1~$\AA, and estimations of the absolute energy scale $E_{\text{unit}}$ and resulting time scale $t_\text{unit}$ can be made according to DFT data. This results in $E_\text{unit}=56$~eV and $t_\text{unit}=11$~fs with the masses set to $m^*_1=1$ and $m^*_2=0.57$ in dimensionless units. Accordingly, we use a timestep of $\Delta t^*=0.01$ (also in dimensionless units), which corresponds to approximately $\Delta t=0.1$~fs. Furthermore, the potential is shifted and truncated to achieve zero potential energy and forces at a cutoff distance $R_c=10~$\AA, i.e.,
\begin{equation}
    V^{\text{trunc}}(r,s_i,s_j) = V(r,s_i,s_j) - V(R_c,s_i,s_j) - \left.\frac{\partial V(r,s_i,s_j)}{\partial r}\right|_{r=R_c}(r-R_c).
\end{equation}

MD calculations were carried out using LAMMPS \cite{LAMMPS} to gather reference data for assessing the accuracy and efficiency of the GPP method. To this end, the above potential was tabulated, and a timestep of $dt^*=0.01$ was used in LAMMPS dimensionless $\texttt{lj}$ units. All computations described in this section were carried out in the (dimensionless) temperature range from $T^*=0$ to $T^*=10^{-3}$, the latter being close to the glass transition temperature of silica, as observed in MD simulations with the described settings. According to \citet{roy2018silica}, $T^*=0.015$ corresponds to approximately $9700$~K, resulting in the highest temperature in this study to be about $650$~K.

\subsection{Thermal expansion}
\label{sec:ThermalExpansion}

\begin{figure}[b!]
\centering
\includegraphics[scale=0.8]{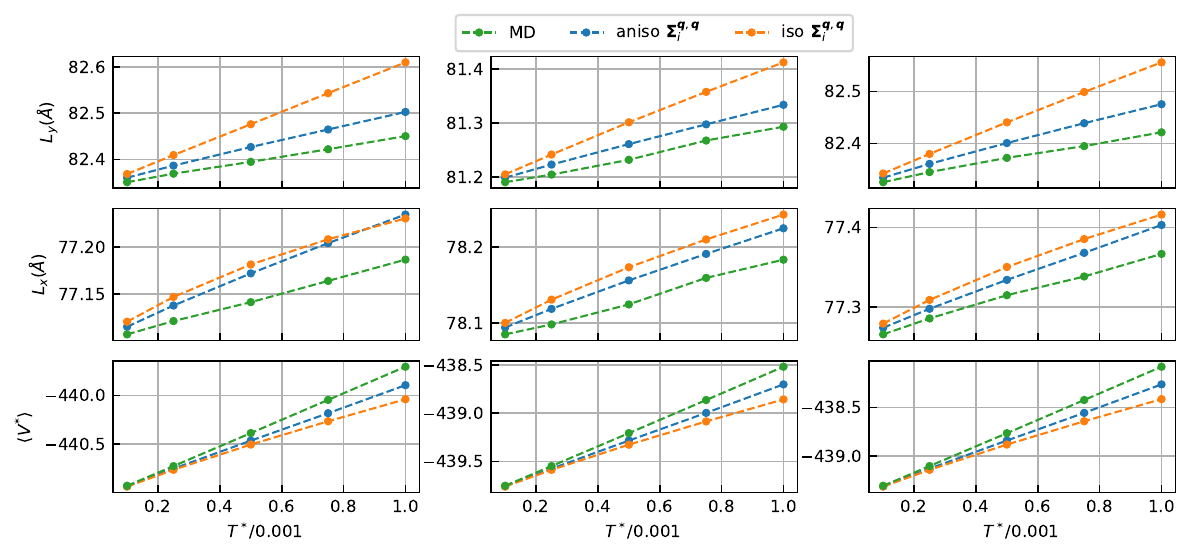}
\caption{Dimensions and average potential energy vs.\ (dimensionless) temperature for the three different 2D silica samples of Figure~\ref{fig:samples_initial} under zero-stress conditions, with increasing disorder (resulting from increasing the number of bond-switching attempts in the Monte-Carlo bond switching algorithm \cite{bamer2023molecular}) from left to right.}
\label{fig_thermal_exp}
\end{figure}

The thermal expansion of samples was simulated by solving Eqs.~\eqref{eq_gpp_iso} and \eqref{eq_gpp_aniso} for the isotropic and anisotropic Gaussian distribution formulations, respectively. The additional constraint of zero thermal stresses is imposed by setting $\sigma_{\alpha\beta}=0$ from Eq.~\eqref{eq_vir_gpp}. Results are shown in Figure~\ref{fig_thermal_exp}, which includes the MD data for comparison. The latter was obtained by equilibrating the sample under NVT conditions and adjusting the box vectors in proportion to the corresponding stress components \citep{cai_molecular_2012} every $10^3$ steps, iteratively until the maximum stress component (averaged over the $10^3$ NVT timesteps) is below $10^{-5}$.

The results show that the GPP model yields accurate predictions for all three samples, each with a different level of disorder. They also illustrate how the error produced by the GPP approximation increases with temperature, reaching approximately $0.2\%$ at the highest temperature for the box vector component $L_y$ and less than $0.1\%$ for both the box vector component $L_x$ and the total average potential energy of the sample, when using the isotropic Gaussian approximation. With the anisotropic Gaussian distribution, the error decreases on average by a factor of approximately two. Overall, the thermally expanded sample dimensions are over-predicted by GPP compared to MD, and the average potential energy is under-predicted. These errors stem from two main sources. The first source is the GPP ansatz for the probability density $f$, which is not necessarily the correct equilibrium probability distribution. Moreover, our GPP ansatz assumes interatomic independence in the covariance matrix, which is a simplification of reality. The second error source is the third-order Gaussian quadrature rule used to approximate the phase-space integrals. For the phase-averaged potential energy, this second error source manifests in two ways. First, the numerical quadrature introduces an error of the phase-averaged stress given some atomic mean positions --- which stems from the error on the phase-averaged forces --- and, consequently, on the sample dimensions when imposing zero stresses. Second, the numerical quadrature introduces an error of the phase-averaged potential energy (which would persist even if the sample dimensions were set to those predicted by MD). Note that these quadrature errors can be reduced while not compromising computational efficiency (in fact, improving it) by tabulating the results of a more accurate numerical integration for the phase averages \cite{li2011diffusive} or by training a machine-learning model \cite{saxena2023gnn}. Given the small errors in Figure~\ref{fig_thermal_exp}, this is not deemed necessary here. In conclusion, this benchmark underlines the effectiveness of the GPP framework to predict the thermal expansion of silica glass and proves the increased accuracy of the newly-introduced anisotropic Gaussian approximation over its isotropic counterpart. The anisotropic Gaussian is therefore the chosen approximation for the remainder of this study.

\subsection{Quasi-static uniaxial loading}

To assess the capabilities of the GPP framework for predicting the onset of inelastic events as a function of temperature, we present the simulation results of uniaxial tensile tests on the same silica samples studied above.
For all samples, the stress-free states resulting from the thermal expansion calculations with the anisotropic Gaussian approximation from Section~\ref{sec:ThermalExpansion} were used as the initial conditions for these simulations. Each of these initial states is defined by the sets of relaxed mean positions, position variances, and box vectors $\overline{\bfq}^0(T)$, $\bfSigma^0(T)$, and $\boldsymbol{H}^0(T)$, respectively. In all simulations, an incremental uniaxial strain is applied to the samples, defined by the strain increments
\begin{equation}
    \Delta\varepsilon = 
    \begin{bmatrix}
    0 & 0 \\ 0 & \frac{\gamma}{L_y}.
    \end{bmatrix}
\end{equation} 
When using the GPP framework (with an anisotropic Gaussian probability density), we solve Eqs.~\eqref{eq_gpp_aniso} in two ways: first, these equations were used as a replacement of $\bff_i=\boldsymbol{0}$ in the numerical realization of the AQS method, which simulates $0$~K conditions. Second, we additionally include rejection sampling through the Metropolis algorithm summarized in Eqs.~\eqref{eq:transition_prob} and \eqref{eq:accep_prob}, where the minimum free energy GPP probability density is used as the trial probability. Both MD and GPP simulation details and results will be summarized in the following.

\subsubsection{Molecular Dynamics}

\begin{figure}[ht!]
    \centering
    \includegraphics[scale=0.8]{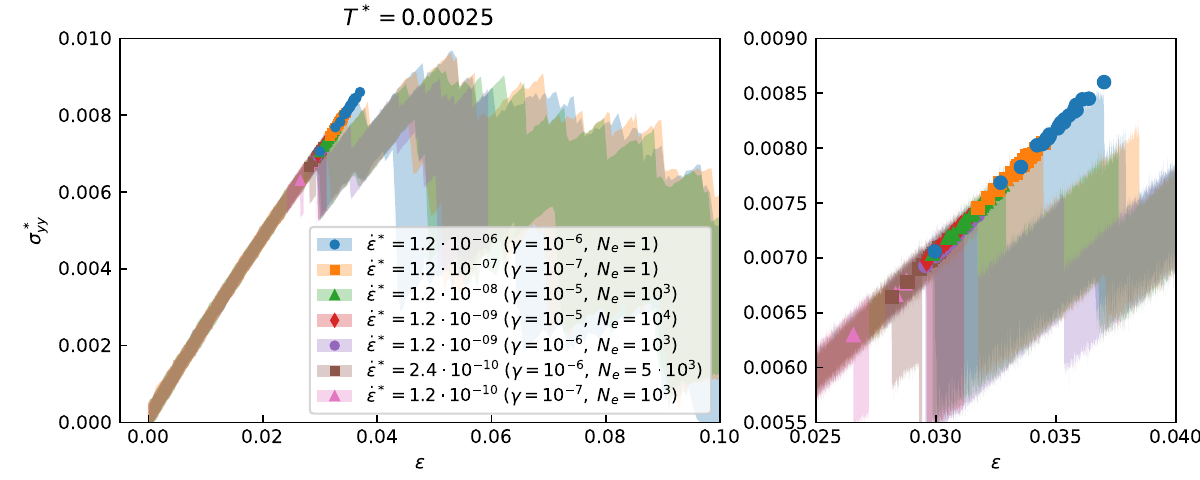}
    \includegraphics[scale=0.8]{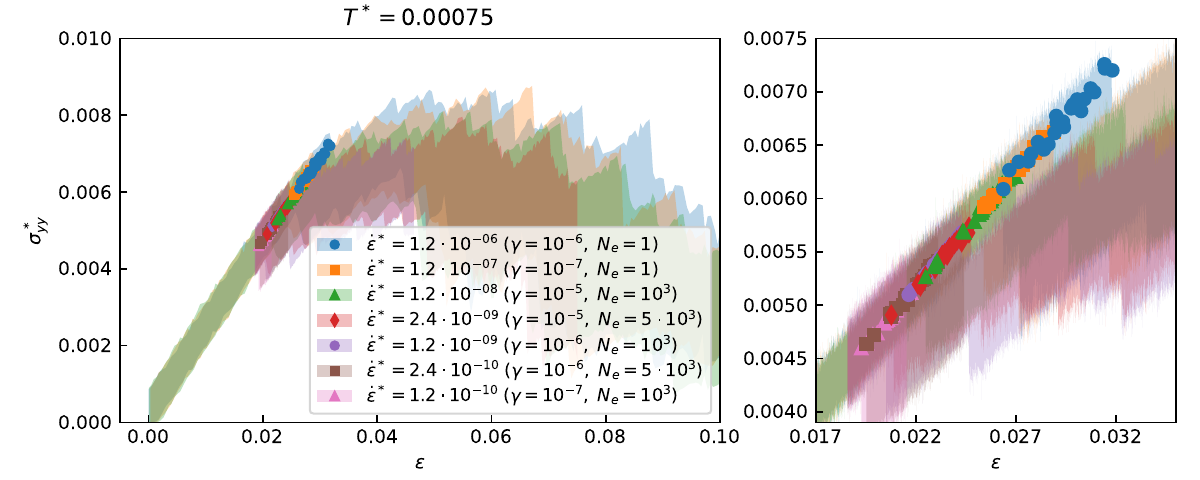}
    \caption{Stress-strain curves obtained from MD simulations for different values of $\{\gamma, N_e\}$ for sample (a) in Figure \ref{fig:samples_initial} at (dimensionless) temperatures $T^*=0.00025$ ($\approx160$~K) and $T^*=0.00075$ ($\approx485$~K). Each shaded region encloses stress-strain curves of the 25 runs with common parameters $\{\gamma, N_e\}$. A marker with matching color is included to denote the $(\varepsilon_{yy},\sigma^*_{yy})$ point at the first inelastic event of each run (where observed). Graphs on the right are magnifications of the onset of inelasticity on the left. For the lower strain rates, samples were pre-strained elastically to avoid long computational times in the elastic regime. Each shaded region covers a strain range achieved by all 25 simulations as not all reached the same final strain.}
    \label{fig:md_uniaxial_by_T}
\end{figure}

Unaxial loading MD simulations were performed by applying the deformation gradient $\boldsymbol{I}+\Delta\varepsilon$ to the instantaneous atomic positions and the simulation box vectors every $N_e$ timesteps, while sampling the NVT ensemble --- effectively increasing the box vector $L_y$ by $\gamma~$\AA~with every deformation increment. To approach quasi-static conditions, in which the system is always in thermodynamic equilibrium, different values of $N_e$ were tested for different selected values of the strain parameter $\gamma$ to allow for equilibration. This allowed to asses if the system was noticeably perturbed out of equilibrium after each application of the deformation. Therefore, different value pairs $\{\gamma,N_e\}$ are used to study their effect on the stress-strain state at which the first inelastic event is observed. This follows previous work in \cite{ochoa_molecular_1991}, which studied the effect of the equilibration of thermal vibrations on the `softening' of the failure process, concluding an equivalence between applying higher $\gamma$-values followed by a number of equilibration time steps and applying lower $\gamma$-values at every timestep. For each value pair, 25 simulations were performed. For the lower strain rates investigated, samples were pre-strained (well within the linear elastic regime prior to the onset of any inelasticity) to save computational time. The results are shown in Figure~\ref{fig:md_uniaxial_by_T} for five different dimensionless strain rates $\dot\varepsilon$. According to the unit time $t_\text{unit}=11fs$ for the settings described above, the lowest strain rate, corresponding to an extension of $\gamma=10^{-7}$~\AA~ applied every $N_e=10^3$ timesteps, is approximately $10^4~s^{-1}$ (see Appendix \ref{sec:strain_rate_eq} for details on the strain rate conversion).

\begin{figure}
    \centering
    \begin{tabular}{cc}
    \includegraphics[scale=0.7]{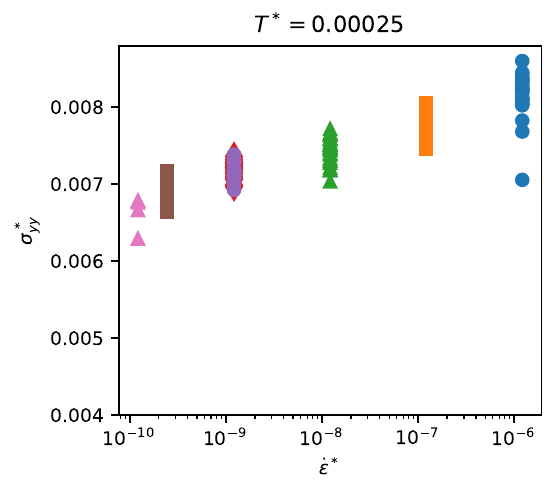} &
    \includegraphics[scale=0.7]{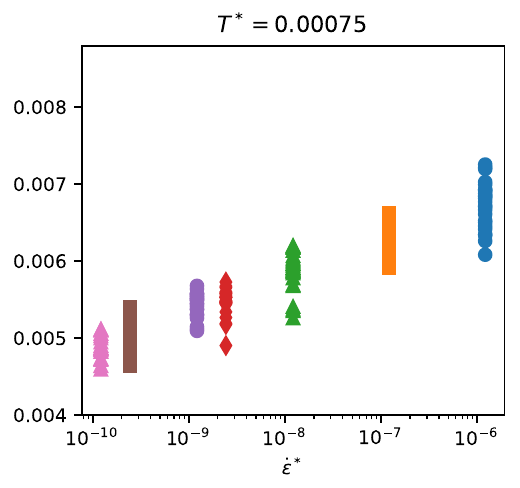}\\
    \hspace{36pt}(a) & \hspace{12pt}(b)
    \end{tabular}
    \caption{Stresses at the onset of inelasticity for the MD simulations at different strain rates. The markers correspond to those in Figure \ref{fig:md_uniaxial_by_T}.}
    \label{fig:md_uniaxial_yields}
\end{figure}

In agreement with previous reports \cite{chowdhury_molecular_2016}, simulations show that lower strain rates generally result in lower stress values at the onset of the first inelastic event. This is clearly shown in Figure~\ref{fig:md_uniaxial_yields}, which summarizes the markers of Figure~\ref{fig:md_uniaxial_by_T} (indicating the onset of inelasticity). It is further evident that the onset of inelasticity decreases with increasing temperature. Variations among the 25 runs for each scenario tend to increase with increasing loading rate, as may be expected.

On a more technical note, different pairs of $\{\gamma, N_e\}$ resulting in equal strain rates $\dot{\varepsilon}$ yield similar results in general. Deviations are visible for some combinations of rate and temperature; e.g., at $T^*=0.00075$ an earlier onset of the first inelastic event is seen at $\dot{\varepsilon}^*=2.4\cdot10^{-9}$ with $N_e=5000$ than at $\dot{\varepsilon}^*=1.2\cdot10^{-9}$ with $N_e=1000$ equilibration steps. Yet, those are minor outliers from the general trend. 
While general conclusions on the effect of different values $\{\gamma, N_e\}$ at the same effective strain rate cannot be drawn from the limited data, the aforementioned observation raises the question if the deformation protocol should be characterized by both parameters instead of only $\dot{\varepsilon}$ (which we will get back to when comparing to results from the Metropolis-GPP method). Even though Metropolis trials and MD steps are not equivalent, in some scenarios a sampling pass can be considered to be proportional to an MD step \cite{bal2014time}. The lowest (dimensionless) strain rate achieved in a reasonable computation time was $\dot{\varepsilon}^*=10^{-10}$ (equivalent to $10^4$~s$^{-1}$, see Appendix~\ref{sec:strain_rate_eq}), which resulted in the lowest stress and strain at the onset of the first inelastic event at all levels of temperature studied. This will be taken as the reference data closest to the quasi-static limit in the following.

\begin{figure}[ht!]
    \centering
    \begin{tabular}{cc}
    \includegraphics[scale=0.65]{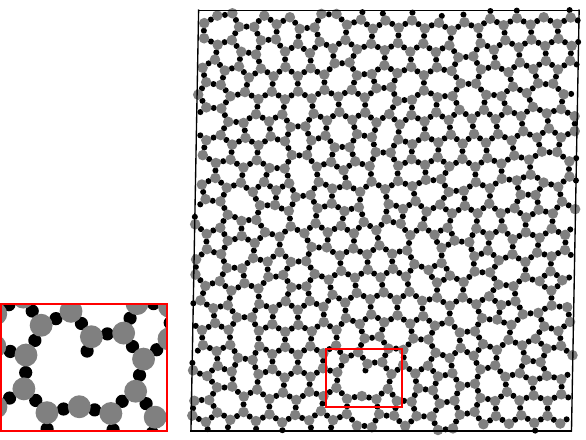} &
    \includegraphics[scale=0.65]{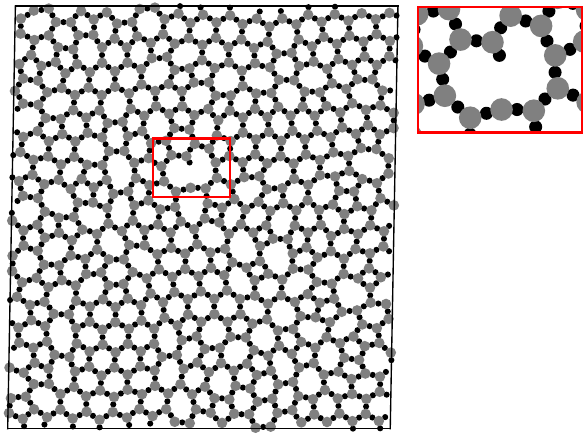} \\
    \hspace{50pt}(a) & \hspace{-50pt}(b)
    \end{tabular}
    \caption{The two observed initial breaking sites for sample (a) of Figure~\ref{fig:samples_initial} in the MD simulations. Breaking site (b) was observed only in a few simulation runs as the first breaking point.}
    \label{fig:md_failure_bonds}
\end{figure}

Finally, we note that the 25 different runs for each case were performed with the goal of sampling potentially different stress-strain responses characteristic of the non-deterministic failure of amorphous solids at finite temperature. The data presented in Figures~\ref{fig:md_uniaxial_by_T} and \ref{fig:md_uniaxial_yields}, which correspond to sample (a) of Figure~\ref{fig:samples_initial}, show two distinct initial failure sites across the 25 runs, which are illustrated in Figure~\ref{fig:md_failure_bonds}. For most of the simulation runs, the failure location of Figure~\ref{fig:md_failure_bonds}(a) was observed. For example, at $T^*=0.00025$ for the case $\{\gamma,N_e\}=\{10^{-6},5\cdot 10^{3}\}$ only in one of the 25 simulations the case of Figure~\ref{fig:md_failure_bonds}(b) was observed; the same is true at $T^*=0.00075$ for the case $\{\gamma,N_e\}=\{10^{-7},10^3\}$ for four of the 25 simulations. The prediction of these initial failure points will be addressed next through the use of the GPP framework. As it will be shown, the position variance $\bfSigma^{(\bfq,\bfq)}$ is a good indicator of potential failure sites for a given initial state. Moreover, with increasing deformation, the latter shows an increasingly strong selection among the possible failure sites.

\subsubsection{Gaussian Phase Packets}
\label{sec:GPPResults}

Uniaxial loading simulations with the GPP framework involve an iterative two-step process, analogous to the AQS method, composed of applying $\Delta\varepsilon$, followed by a relaxation of the mean positions $\overline{\bfq}$ and position variances $\bfSigma^{(\bfq,\bfq)}$. More specifically, at iteration $n$, the uniaxial deformation maps each relaxed atomic mean position $\overline\bfq_i^{n-1}$ and the box vectors $\boldsymbol{H}^{n-1}$ from the previous iteration to a new, unrelaxed position $\bfq_i^{*n}$ and updated box vectors $\boldsymbol{H}^n$, i.e.,
\begin{equation}
\bfq_i^{*n} = (\boldsymbol{I}+\Delta\varepsilon) \overline{\bfq}_i^{n-1}\qquad\text{and}\qquad\boldsymbol{H}^{n} = (\boldsymbol{I}+\Delta\varepsilon) \boldsymbol{H}^{n-1}.
\end{equation}%
This set of deformed positions $\bfq^{*n}$ and the set of relaxed position variances from the previous iteration $\bfSigma^{n-1}$ are used as initial guess to solve Eqs.~\ref{eq_gpp_aniso}, while the simulation box is fixed to maintain the applied stress, resulting in $\overline{\bfq}^n$ and $\bfSigma^n$. This process is repeated for each load increment at a given temperature.

\begin{figure}[t!]
    \centering
    \includegraphics[scale=0.8]{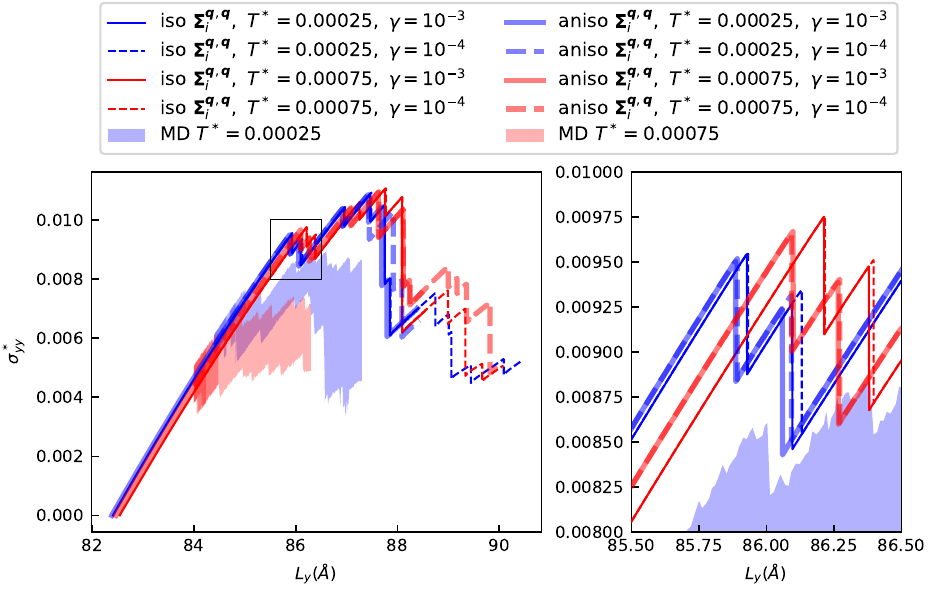}
    \caption{Stress-strain curves and a magnification of the section of the onset of inelasticity (highlighted on the left) for sample (a) of Figure \ref{fig:samples_initial} at $T^*=0.00025$ and $T^*=0.00075$, using the isotropic and anisotropic Gaussian approximations. Shaded regions depict the three cases with lowest strain rates for each temperature of the MD results shown in Figure~\ref{fig:md_uniaxial_by_T}.}
    \label{fig:no_mc}
\end{figure}

\begin{figure}[ht!]
    \centering
    \begin{tabular}{c}
    \includegraphics[scale=0.55]{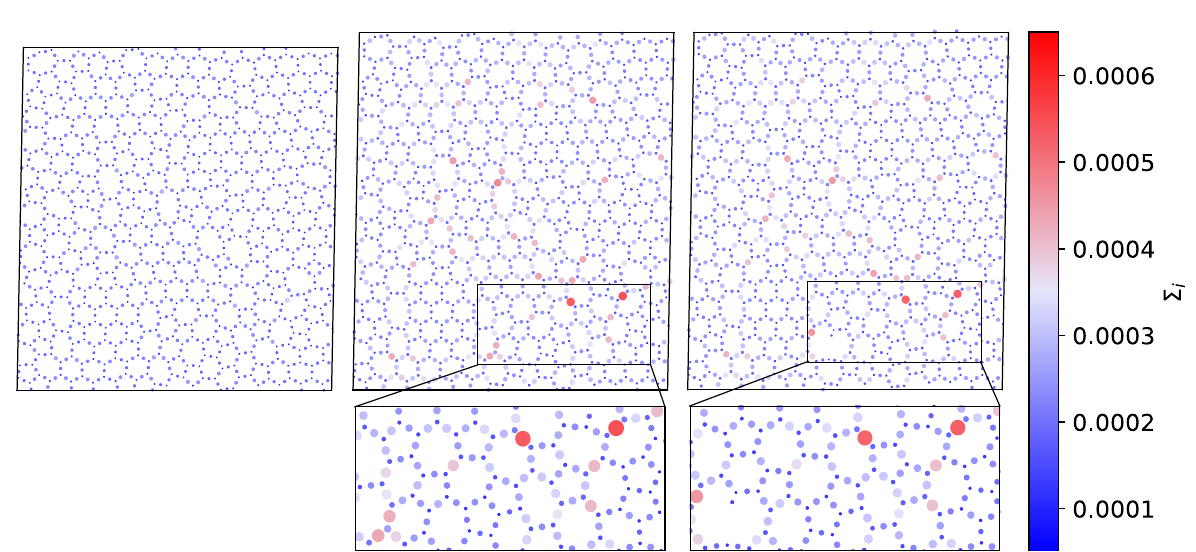} \\
    (a)~isotropic Gaussian \\
    \includegraphics[scale=0.55]{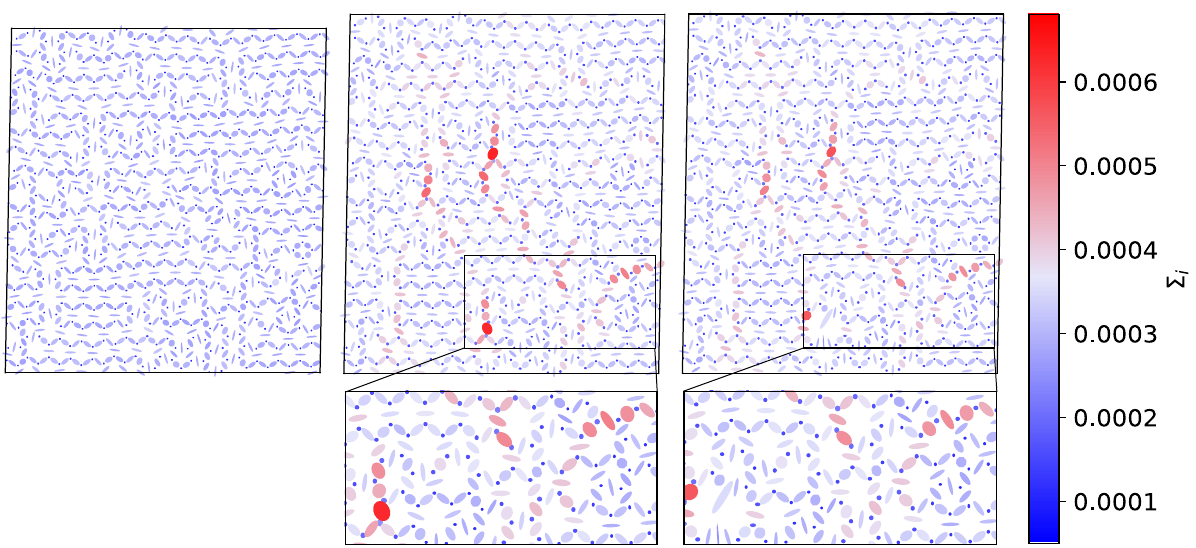}\\
    (b)~anisotropic Gaussian
    \end{tabular}
    \caption{Different stages during the uniaxial deformation response shown of the cases corresponding to $T^*=0.00025$ and $\gamma=10^{-3}$ in Figure~\ref{fig:no_mc} using (a) the isotropic and (b) the anisotropic Gaussian approximation. For each row, shown are the initial state (left), the state before the first inelastic event (middle), and the state after first inelastic event (right). The color of an atom $i$ corresponds to the value $\Sigma_i^{2} = \text{det}\bfSigma^{(\bfq,\bfq)}_i$. Principal axes of the ellipses representing the atoms in the anisotropic case are aligned with the eigenvectors of $\bfSigma_i^{(\bfq,\bfq)}$ and re-scaled.}
    \label{fig:no_mc_sequence}
\end{figure}

Solving the quasi-static GPP equations implies finding a static probability density, which is why there is inherently no time dependence in this process, and a finite strain cannot be defined. Furthermore, the size of $\gamma$ sets the initial guess of the mean atomic positions for the solution of Eqs.~\ref{eq_gpp_aniso} and thus affects the sequence of free energy minima that the system will access during a simulation, as discussed in Section~2. Contrary to MD, the initial guess here determines the accessible free energy minima of the ensemble, as the relaxation process ends at a static ensemble described by $f$ with minimum free energy (in comparison, the relaxation in the AQS ends at a static microstate with minimum potential energy). Therefore, investigating the influence of the size of the applied $\Delta\varepsilon$ is crucial, which is why different values were investigated. The results of these simulations are presented in Figure~\ref{fig:no_mc}, and the deformed configuration of the corresponding sample at different stages is shown in Figure~\ref{fig:no_mc_sequence}. Only a single run is performed, as this methodology results in repeatable results in a deterministic fashion, since the relaxation of the ensemble results in the same free energy minimum for given applied strain and initial phase-averaged mechanical and thermal forces.

The effect of the thermal expansion of the sample (see Section~\ref{sec:ThermalExpansion}) can also be observed in Figure~\ref{fig:no_mc}, as the initial dimensions differ at different temperatures. As already observed above, the anisotropic Gaussian approximation shows reduced thermal expansion --- closer to the MD calculations --- in comparison to the isotropic Gaussian. However, the effect of temperature on the stress and strain values at the onset of the first inelastic event is not in line with the above MD data. In MD (and in physical reality), the effect of temperature on driving systems across potential energy barriers of inelastic events enters through both the thermally expanded state and the thermal fluctuations about that mean state. While the changes due to thermal expansion in the GPP procedure are accounted for, the effect of thermal fluctuations needed for such transitions as observed in the MD results is not captured. It is therefore necessary to extend the method if the effect of these thermal fluctuations is to be captured, which will be accomplished through the Markovian process described in Section~\ref{sec:ResultsMonteCarlo}. 

Irrespective of the accuracy of the stress-strain curves, one important observation is that the anisotropic Gaussian parameters $\{\bfSigma^{(\bfq,\bfq)}_i\}$ provide an accurate estimator of sites for initial inelastic events (with higher accuracy than the isotropic Gaussian). As observed in Figure~\ref{fig:no_mc_sequence}(b), the anisotropic Gaussian approximation clearly shows how, before the first inelastic event (middle row), there are two oxygen atoms with the highest $\Sigma_i = \left(\text{det}\bfSigma^{(\bfq,\bfq)}_i\right)^{1/2}$ values, which are exactly the sites where the first inelastic events were observed in the MD calculations (cf.~Figure~\ref{fig:md_failure_bonds}). The four zoomed-in regions of Figure~\ref{fig:no_mc_sequence} show the sites where the first bond breaks and surrounding atoms with high $\Sigma$ values. While the isotropic Gaussian case in Figure~\ref{fig:no_mc_sequence}(a) predicts atoms with higher $\Sigma$ at sites different from where the bond breaks, the anisotropic case in Figure~\ref{fig:no_mc_sequence}(b) yields the highest $\Sigma$ exactly for those two sites where the MD calculations showed the first broken bonds. 

\begin{figure}
    \centering
    \includegraphics[scale=0.7]{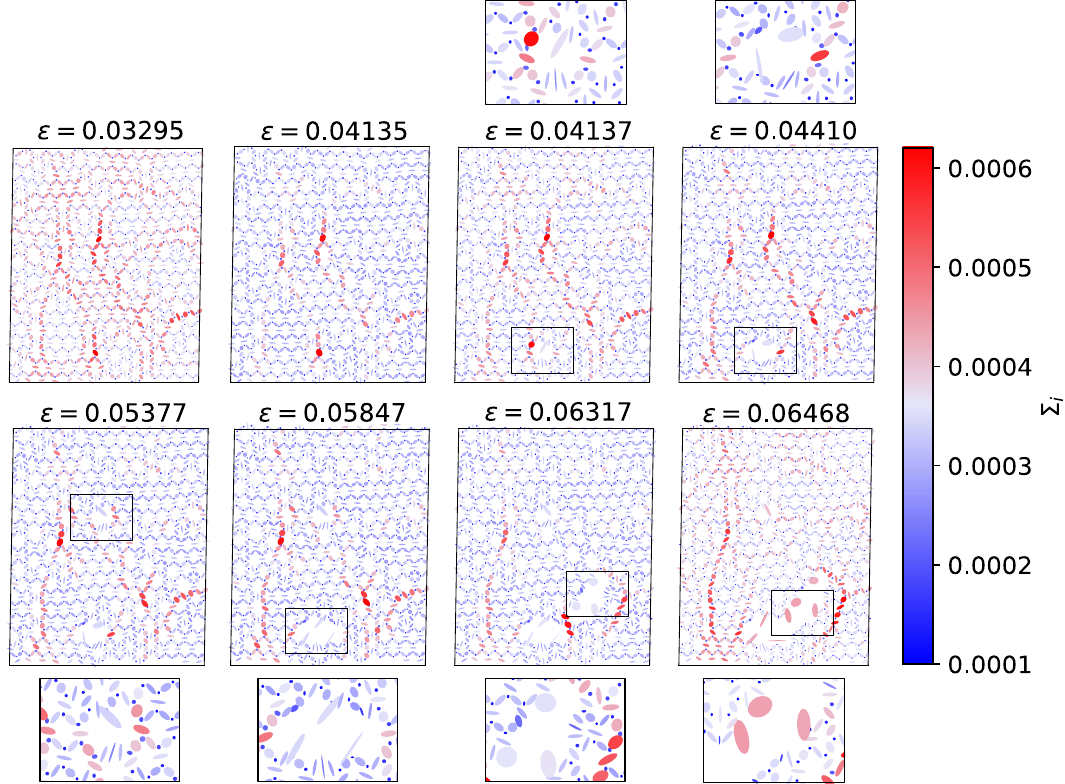}
    \caption{Snapshots at different strains of sample (a) in Figure~\ref{fig:samples_initial} for temperature $T^*=0.00025$ and loading $\gamma=10^{-4}$, using the anisotropic Gaussian approximation. One snapshot is shown for every time a bond failure is detected, except for the first two snapshots, where all bonds are still intact. Black rectangles highlight those regions with detected broken bonds, which are also shown with magnification above (top row) and below (bottom row). The color of an atom $i$ corresponds to the value $\Sigma_i^{2} = \det\bfSigma^{(\bfq,\bfq)}_i$. Principal axes of the ellipses representing the atoms in the anisotropic case are aligned with the eigenvectors of $\bfSigma_i^{(\bfq,\bfq)}$ and re-scaled.}
    \label{fig:no_mc_sequence_long}
\end{figure}

A closer inspection of the failure progress is shown in Figure~\ref{fig:no_mc_sequence_long} for the same sample (a) of Figure~\ref{fig:samples_initial} at increasing values of strain. One snapshot is recorded for every time a bond failure is detected, except for the first two snapshots, where all bonds are intact. The broken bond at each stage is marked by a black open square with a corresponding magnified view above (top row) or below (bottom row). It becomes apparent that the broken bond in any snapshot is indeed the bond formed by an atom with the first- or second-highest $\Sigma$-value in the previous snapshot (before the bond is broken). Therefore, the anisotropic Gaussian approximation cannot only be viewed as a predictor of the potential failure points in the initial state of the sample (first snapshot) but also throughout the entire deformation protocol. As the observed failure points are dependent on the load history and, in general, cannot be determined a priori from the reference state alone, it is beneficial that this method tracks how the failure probability of each site evolves as the deformation progresses.

\subsubsection{Metropolis-Gaussian Phase Packets}
\label{sec:ResultsMonteCarlo}

The above results demonstrated that the GPP framework, solving Eqs.~\ref{eq_gpp_aniso} after applying strain increments $\Delta\varepsilon$ to the atomic positions $\overline\bfq$, does not predict transitions in the free energy landscape with the same temperature dependence as in MD. In order to predict such transition events, associated with PES transitions at lower stresses due to the effect of thermal fluctuations, the Metropolis algorithm discussed in Section~\ref{sec:MC} is introduced into GPP simulations. The resulting extended procedure consists of performing the Metropolis algorithm after every GPP relaxation (i.e., at constant strain), which allows to better sample the free energy landscape. This admits reaching new starting guesses for the relaxed state in the neighborhood of potentially nearby free energy minima. After every deformation increment, the system returns to a free energy minimum by solving Eqs.~\ref{eq_gpp_aniso}, thus approaching quasi-static conditions, but at the same time introducing an effect mimicking that of thermal fluctuations and potentially triggering transitions. An advantage of this extension is that the relaxed probability density is a natural choice for the trial probability. It is common practice in molecular simulations involving rejection sampling methods to adjust the parameters of the trial probability with the goal of attaining an optimal acceptance rate \cite{tuckerman2023statistical}. In our procedure, such an adjustment is not necessary (though it is possible and can be informed by the GPP probability density if desired), yielding acceptance rates for the examples in this work in the range of $60-70$\%.

\begin{figure}[ht!]
    \centering
    \includegraphics[scale=0.7]{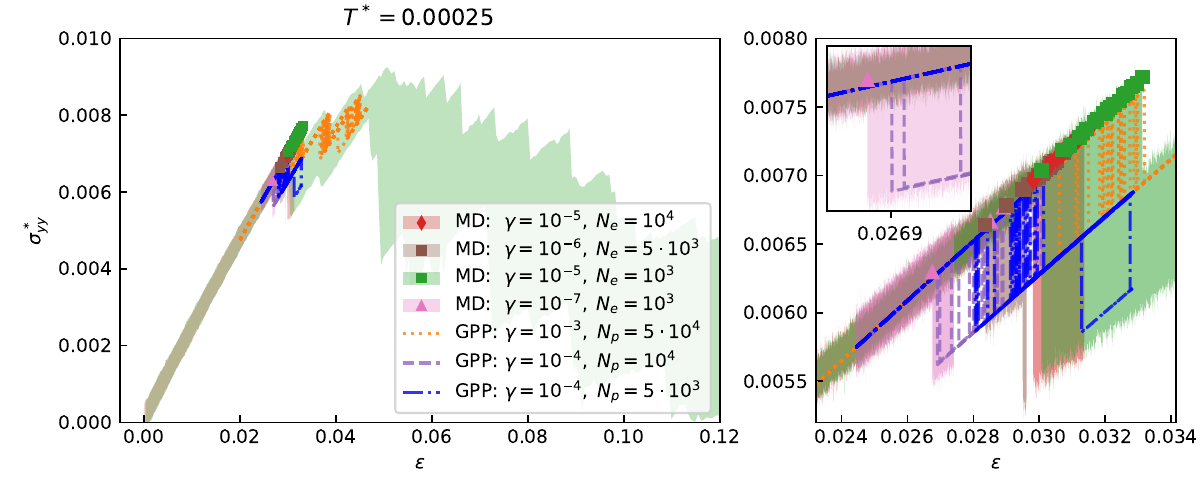}
    \includegraphics[scale=0.7]{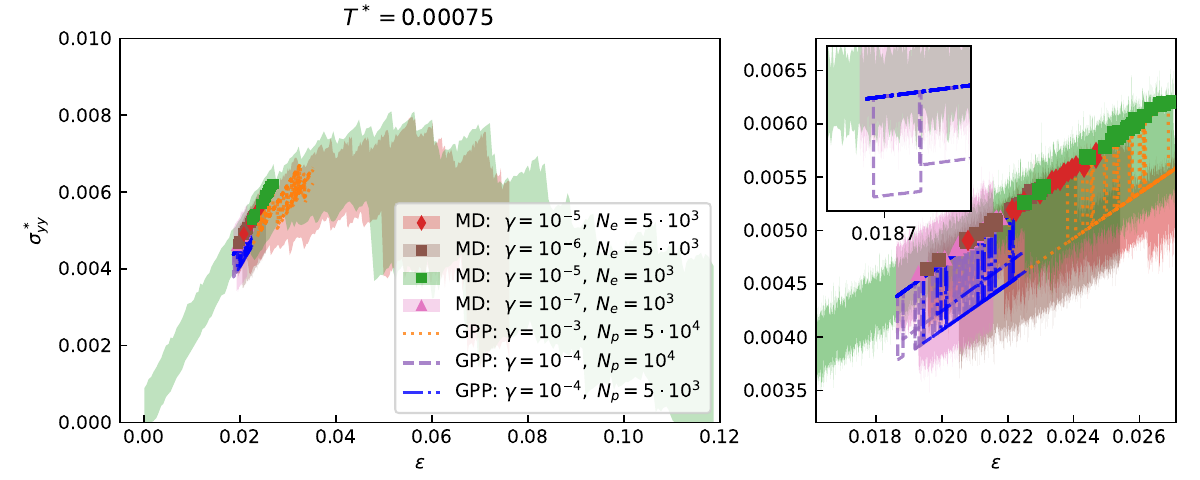}
    \caption{Stress-strain responses obtained from the Metropolis-GPP framework, using the anisotropic Gaussian approximation, compared to selected MD results (from Figure~\ref{fig:md_uniaxial_by_T}) including the three lowest strain rates at dimensionless temperatures $T^*=0.00025$ (top) and $T^*=0.00075$ (bottom). Shown on the right are magnifications of the onset of inelasticity in the full stress-strain curves on the left. Analogous to the MD calculations, Metropolis-GPP samples were pre-strained during the elastic regime to reduce computational costs.}
    \label{fig:gpp-mc_vs_md}
\end{figure}

With this new procedure, random initial guesses for $\overline\bfq^0$ and $\bfSigma^0$ are produced prior to every GPP relaxation. This introduces randomness in the stress-strain behavior of the system, resulting in an onset of inelastic events at potentially different strain levels during different runs, given the same initial zero-stress state. By contrast, the procedure without Metropolis sampling (see Section~\ref{sec:GPPResults}) always deterministically produces the same stress-strain path for a given initial zero-stress condition without sampling the free energy landscape. Here, the extended procedure is capable of capturing a non-deterministic sequence of visited free energy minima associated with different failure paths. As a consequence, 25 simulation runs (starting from the same initial zero-stress state) were performed, analogous to the MD case, with the goal of reproducing the statistics of the MD procedure more efficiently by the GPP framework. A significant gain in efficiency can be expected if the Metropolis-GPP scheme reproduces the mean thermal softening and onset of inelasticity with larger values of $\gamma$ than MD. Higher $\gamma$-values require a lower number of incremental deformation steps to achieve a total given strain, resulting in faster simulations. Of course, this assumes that the computational cost of solving the GPP quasi-static equations plus sampling $N_s$ trials at each of these incremental deformation steps is less expensive (or at least not considerably more expensive) than running MD between incremental deformation steps. To investigate this, different combinations of $\gamma$ and the number of Metropolis trials are studied. The results are presented in Figure~\ref{fig:gpp-mc_vs_md}.

\begin{figure}
    \centering
    \includegraphics[scale=0.7]{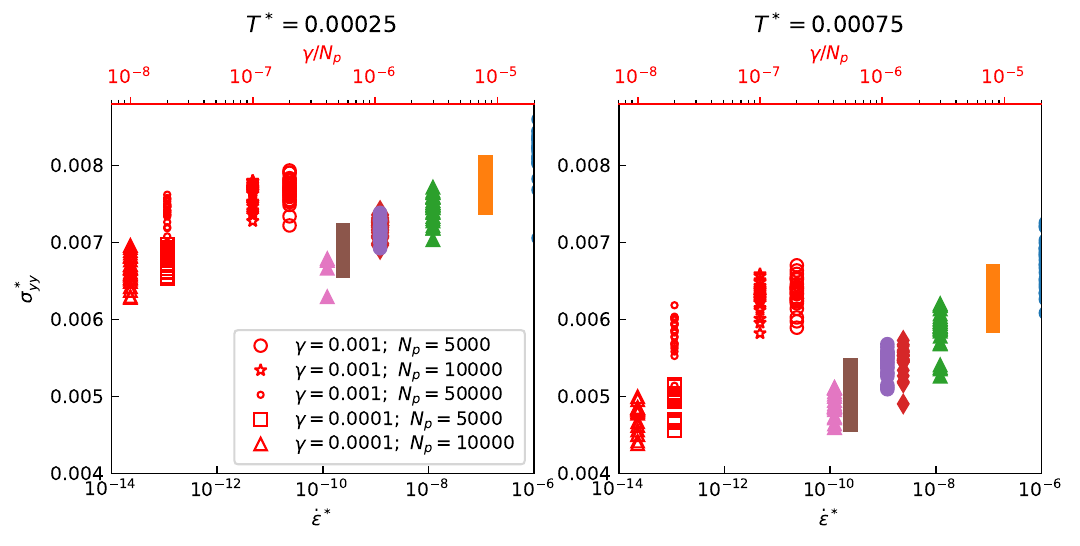}
\caption{Stresses at the onset of inelasticity in simulations using the Metropolis-GPP framework (red) as a function of the ratio between the incremental deformation parameter $\gamma$ and the number of Metropolis passes $N_p$ (top axis). For comparison, the stresses at the onset of inelasticity in the MD simulations from Figures~\ref{fig:md_uniaxial_by_T} and~\ref{fig:md_uniaxial_yields} are also shown as a function of the (dimensionless) strain rate $\dot\varepsilon^*$ (bottom axis).}
    \label{fig:gpp-mc_vs_md_yields}
\end{figure}

These results confirm that larger values for the deformation increment $\gamma$ in Metropolis-GPP can yield inelastic onsets as low as in MD for a given temperature, accurately reflecting the temperature dependence. This is, however, only true for sufficiently many Metropolis trials. Specifically, at both temperatures shown in Figure~\ref{fig:gpp-mc_vs_md} the first inelastic onset in the different MD runs with $\dot{\varepsilon}^*=2.4\cdot10^{-10}(\gamma=10^{-6}, N_e=5\cdot10^3)$ and $\dot{\varepsilon}^*=1.2\cdot10^{-10}(\gamma=10^{-7}, N_e=5\cdot10^3)$ are close to those predicted by the Metropolis-GPP procedure using $\{\gamma,N_p\}=\{10^{-4},5\cdot10^{3}\}$ and $\{\gamma,N_p\}=\{10^{-4},10^{4}\}$, respectively. This comparison is more clearly visualized in Figure~\ref{fig:gpp-mc_vs_md_yields}. The overall efficiency gain rests on the combination of the individual efficiency gain from taking larger deformation increments $\gamma$ and the efficiency gain/loss of performing relaxation with Metropolis sampling vs.\ integrating the MD equations of motion over $N_e$ time steps.

The efficiency gain dependent on $\gamma$ is limited by the potential energy landscape of the system, as it also defines the free energy landscape given a specific $f$. For large increments $\gamma$, the ensemble will be mapped to a point in the free energy surface where intermediate free energy minima between this state and the minimum free energy state prior to the deformation increment are no longer reachable by a relaxation towards Eq.~\eqref{eq_gpp_aniso}. At the same time, the efficiency of the relaxation towards Eq.~\eqref{eq_gpp_aniso} can be improved, e.g., by more efficient quadrature methods (see Section~\ref{sec:ThermalExpansion}). Additionally, the convergence of the relaxation may be sped up by optimizing the FIRE parameters or resorting to other gradient-based solvers. For improving the efficiency of the sampling, parallelization schemes; see, e.g., \cite{Calderhead2014parallelMetropolis}. It is worth noting that sampling methods like the Metropolis algorithm also admit parallelization across replica, providing more efficient statistical studies than when using MD. In this work, each MD simulation was performed running LAMMPS on eight cores, while each Metropolis-GPP simulation was performed on one core of the same hardware. To give a sense of the computational cost of GPP in the current implementation (which has not been optimized for high-performance computing), an MD simulation with $\dot{\varepsilon}^*=2.4\cdot10^{-10}(\gamma=10^{-6},N_e=5\cdot10^3)$ and a GPP simulation with $\{\gamma,N_p\}=\{10^{-4},5\cdot10^{3}\}$ achieved a total extension of the simulation box component $L_y$ of about $0.76$\AA~and $0.70$\AA, respectively, from the same initial state in the same wall-clock time. This may seem to reflect that the GPP procedure does not result in important gains. However, it must be reiterated that the GPP simulations were not parallelized and hence run on a single core (vs.\ eight cores in the MD simulations) and that its current implementation offers significant opportunity for algorithmic optimization, including more efficient numerical integration in phase space, neighbor list updates, solvers, etc. (which are highly optimized in LAMMPS). More importantly, the anisotropic Gaussian covariance matrix $\bfSigma^{\bfq,\bfq}_i$ for each atom, which has been shown above to be an accurate predictor of failure sites, is directly available for every relaxed state along the loading history. This is a clear advantage over MD, where calculating this covariance matrix would significantly increase the computational cost (incurred by the calculation steps for taking time averages of $\bfSigma^{\bfq,\bfq}_i$ for every single atom and indirectly by the need for longer MD simulations between loading increments to ensure statistical convergence of this covariance parameters).

Two more observations of the above results are worth noting. First, in most of the Metropolis-GPP simulations shown in Figure~\ref{fig:gpp-mc_vs_md}, the initial inelastic event observed was that of Figure~\ref{fig:md_failure_bonds}(a). For example, at $T^*=0.00025$ the case in Figure~\ref{fig:md_failure_bonds}(a) was observed in one and three of the 25 simulations for the cases $\{\gamma, N_p\}=\{10^{-4},5\cdot10^3\}$ and $\{\gamma, N_p\}=\{10^{-4},10^4\}$, respectively. This is in line with the MD results discussed above. Second, for $\gamma=10^{-4}$, the Metropolis-GPP results show rebonding in some of the replica. This can be seen, e.g., in Figure~\ref{fig:gpp-mc_vs_md} (bottom right) for the replica with the earliest inelastic onset among the shown simulations at a strain of about $\varepsilon=0.0187$ with $\{\gamma,N_p\}=\{10^{-4},10^4\}$ (violet line, more clearly shown in the magnified view showing the three (for $T^*=0.00025$) and two (for $T^*=0.00075$) earliest stress drops in the GPP simulations). The stress decreases initially and then increases at a slightly higher strain, which corresponds to a Si-O pair in the zoomed-in area in Figure~\ref{fig:md_failure_bonds}(a) detaching and re-bonding. This showcases the statistical characteristics of the GPP framework, whose phase-averaging across different microstates determines the relaxation of the initial guess $\{\overline\bfq^0,\bfSigma^0\}$ through $\langle\bff\rangle$. In this case, the Metropolis scheme samples initial guesses, where $\langle\bff\rangle$ has contributions of microstates with and without the bond broken (located near two different corresponding PES minima). Depending on $\overline\bfq$ and $\bfSigma$ of the initial guess, those microstates with the bond broken or those with the bond intact will dominate $\langle\bff\rangle$ and the relaxation will proceed accordingly towards one of the corresponding equilibrium states.

\section{Conclusion}
\label{sec:conclusions}

We have extended the finite-temperature atomistic modeling framework based on Gaussian Phase Packets (GPP) to disordered solids, as demonstrated by the application to 2D silica glasses under mechanical loading at finite temperature --- as a potential avenue for more efficient atomistic calculations than with MD while allowing for the prediction of rearrangements sites in the form of local fracture events, which can be seen equivalent to shear transformation zones (STZs) in sheared metallic glasses, thus extending the commonly used Athermal Quasi-Static method from zero to finite temperature. We have applied this framework to silica samples to compute the thermal expansion, the location of rearrangement sites, and the effect of temperature on the onset of the inelastic regime at low strain rates. The GPP framework, with the newly introduced anisotropic Gaussian approximation, provides good estimates of thermal expansion (compared to MD) and is also a good predictor of local rearrangement sites, both when applied to undeformed samples and, with improved accuracy, to deformed samples as deformation progresses. Furthermore, combined with the presented Metropolis algorithm informed by the GPP PDF, predictions at different temperatures match those of MD for the onset of inelasticy at the lowest studied strain rates. For such matching cases, we compare the ratios of the incremental deformation parameter $\gamma$ and the number of integration steps between deformation steps in MD ($\gamma/N_e$) and the number of Metropolis sampling passes in Metropolis-GPP between deformation steps ($\gamma/N_s$). This gives an idea of efficiency gains (when assuming each integration step and each Metropolis pass are roughly equally expensive), which yields that the Metropolis-GPP ratio can handle ratios $\gamma/N_s$ that are about two orders of magnitude larger than $\gamma/N_e$ in MD.

The efficiency gain of the Metropolis-GPP framework, one of its potential advantages, was not considerable in this study due to the low efficiency of the current Gaussian quadrature implementation for computing phase space averages. The computational complexity of the algorithm allows it to be significantly optimized, which will be addressed in the future. Furthermore, we emphasize that this framework also yields individual atomic frequencies encoded in the Gaussian covariance matrix, which has proven to be a good predictor of rearrangement zones, even if the interatomic independence destroys the full vibrational footprint. More importantly, it scales better than computing full dynamical matrices or even individual atomic position covariances through time averages during MD simulations to obtain equivalent predictions of rearrangement sites.

The efficiency gain of the Metropolis-GPP framework is also limited by the required number of Metropolis passes to ensure an extensive sampling of initial ensemble conditions for relaxing towards relevant adjacent free energy minima. Improved sampling techniques in combination with other methods, such as methods for thermalizing the potential energy landscape resulting in lower free energy barriers, will be further explored, as this can greatly improve the overall efficiency of the framework.

Finally, other systems, ranging from crystalline to three-dimensional amorphous solids, should be investigated within the presented framework to quantify the improvement of the anisotropic Gaussian approximation over the isotropic case for different systems.

Overall, we have presented a new framework, which serves as a good and inexpensive predictor of local rearrangement sites (analogous to STZs in sheared metallic glasses) at the undeformed state and which performs increasingly better with increased deformation, while, at the same time, having the potential to improve considerably the efficiency of low-strain-rate MD simulations for studying the finite-temperature mechanical properties of solids.

\section*{Acknowledgment}
The support from the European Research Council (ERC) under the European Union’s Horizon 2020 research and innovation program (grant agreement no.~770754) is gratefully acknowledged.

\appendix

\section{Strain rate non-dimensionalization} \label{sec:strain_rate_eq}

In this work, the constant unit cell extension $\gamma$ during the deformation increments results in constant engineering strain rates following other MD works such as \citep{dupont_strain_2012,chowdhury_molecular_2016}. The resulting strain rate $\dot\varepsilon^*$ is the engineering strain applied between the beginning of the simulation ($t_0$) and some time $t$ divided by $\Delta t=t-t_0$:
\begin{equation}
    \dot\varepsilon = \frac{\frac{L_y(t)-L_y(t_0)}{L_y(t_0)}}{t-t_0}.
\end{equation}
The case with the lowest strain rate corresponds to an extension of $\gamma=10^{-7}~$\AA~every $N_e=10^3$ timesteps of size $dt^*=0.01$ ($(\cdot)^*$ denotes dimensionless units). These dimensionless times are normalized by a time unit of $t_\text{unit}=11$~fs, as explained in Section \ref{sec:results}. For the sample corresponding to Figure \ref{fig:md_uniaxial_by_T}, the initial value of the box vector component $L_y$ was approximately $L_y(t_0)=82.4~$\AA~in the range of temperatures studied. Thus, this results in a (dimensionless) strain rate of
\begin{equation}
    \dot\varepsilon^* = \frac{\frac{\gamma}{L_y(t_0)}}{N_e\cdot dt^*}=\frac{\frac{10^{-7}\text{\AA}}{82.4\text{\AA}}}{10^3\cdot 0.01}\approx 1.21\cdot10^{-10}
\end{equation}
in reduced time units. With $t_{\text{unit}}=11$~fs, this results in the strain rate
\begin{equation}
\dot\varepsilon=\dot\varepsilon^*\frac{10^{15}\text{fs}/\text{s}}{11\text{fs}}\approx 1.1\cdot 10^{4}\text{s}^{-1}.
\end{equation}

\bibliographystyle{cas-model2-names}
\bibliography{lit}

\end{document}